\documentclass[aps,prl,twocolumn,superscriptaddress,footinbib]{revtex4-2}

\usepackage{amsmath}
\usepackage{amssymb}
\usepackage{graphicx}
\usepackage{xcolor}
\usepackage{hyperref}

\newcommand{\spara}{\scriptscriptstyle\parallel}
\newcommand{\sperp}{\scriptscriptstyle\perp}
\newcommand{\snum}[1]{\scriptscriptstyle (#1)}

\begin{document}

\title{Electric-Field Induced Spin Wave Nonreciprocity in Noncoplanar Magnets}

\date{\today}

\author{Xiao-Hui Li}
\affiliation{Institute of Physics, Chinese Academy of Sciences, Beijing 100190, China}
\affiliation{University of Chinese Academy of Sciences, Beijing 100049, China}

\author{Yuan-Ming Lu}
\email{lu.1435@osu.edu}
\affiliation{Department of Physics, Ohio State University, Columbus OH 43210, United States}
\affiliation{Institute for Solid State Physics, University of Tokyo, Kashiwa 277-8581, Japan}

\author{Yuan Wan}
\email{yuan.wan@iphy.ac.cn}
\affiliation{Institute of Physics, Chinese Academy of Sciences, Beijing 100190, China}
\affiliation{Songshan Lake Materials Laboratory, Dongguan, Guangdong 523808, China}

\begin{abstract}
We show that an electric field can induce nonreciprocal spin wave dispersion in magnetic insulators with negligible spin-orbit coupling. The electric field controls the direction and magnitude of nonreciprocity through a nonlinear magnetoelectric effect without switching the magnetic ground state. By deriving spin space group symmetry constraints, we find only a subset of noncoplanar magnets exhibits this property, and identify a few candidates. For the example of hexagonal lattice tetrahedral antiferromagnet, our effective field theory analysis and microscopic model calculation yield results that are fully consistent with the symmetry analysis.
\end{abstract}

\maketitle

A wave exhibits \emph{nonreciprocity} when its transmission in one direction is different from the opposite direction~\cite{Sounas2017,Nassar2020}. This unusual phenomenon holds far-reaching technological implications exemplified by optical isolators~\cite{Jalas2013} and acoustic rectifiers~\cite{Liang2010}. For spin waves, seeking efficient ways to generate and control their nonreciprocity~\cite{Ishibashi2020,Kanj2023,Guckelhorn2023} is essential for developing magnon logic devices~\cite{Jamali2013,Lan2015,Chumak2015,Cheng2016}.

Nonreciprocal wave dispersion, namely $\omega(q) \neq \omega(-q)$, requires breaking both time reversal and inversion symmetry as either symmetry alone protects the reciprocity. In magnets, time reversal symmetry is broken spontaneously. The inversion symmetry may be broken in several ways. It can be broken by the crystal lattice~\cite{Udvardi2009,Gitgeatpong2017} or magnetic order~\cite{Takuya2020,Ogawa2021,Go2022} in non-centrosymmetric magnets~\cite{Sato2019}, by the electric current in conducting ferromagnets~\cite{Vlaminck2008}, or by the boundary in the case of surface spin waves~\cite{Camley1987,McClarty2022}.

In this work, we expand this toolbox by showing that an electric field can give rise to nonreciprocal spin wave dispersion in an insulating magnet. As the electric field is the sole cause of inversion symmetry breaking, it dictates the nonreciprocity's direction and magnitude. Crucially, the mechanism does not require SOC or switching the ground state, thereby setting it apart from that of non-centrosymmetric magnets~\cite{Sato2019}. It is similar in spirit to the current-induced nonreciprocity~\cite{Vlaminck2008} but mitigates Joule heating.

\begin{figure}
\centering
\includegraphics[width  = \columnwidth]{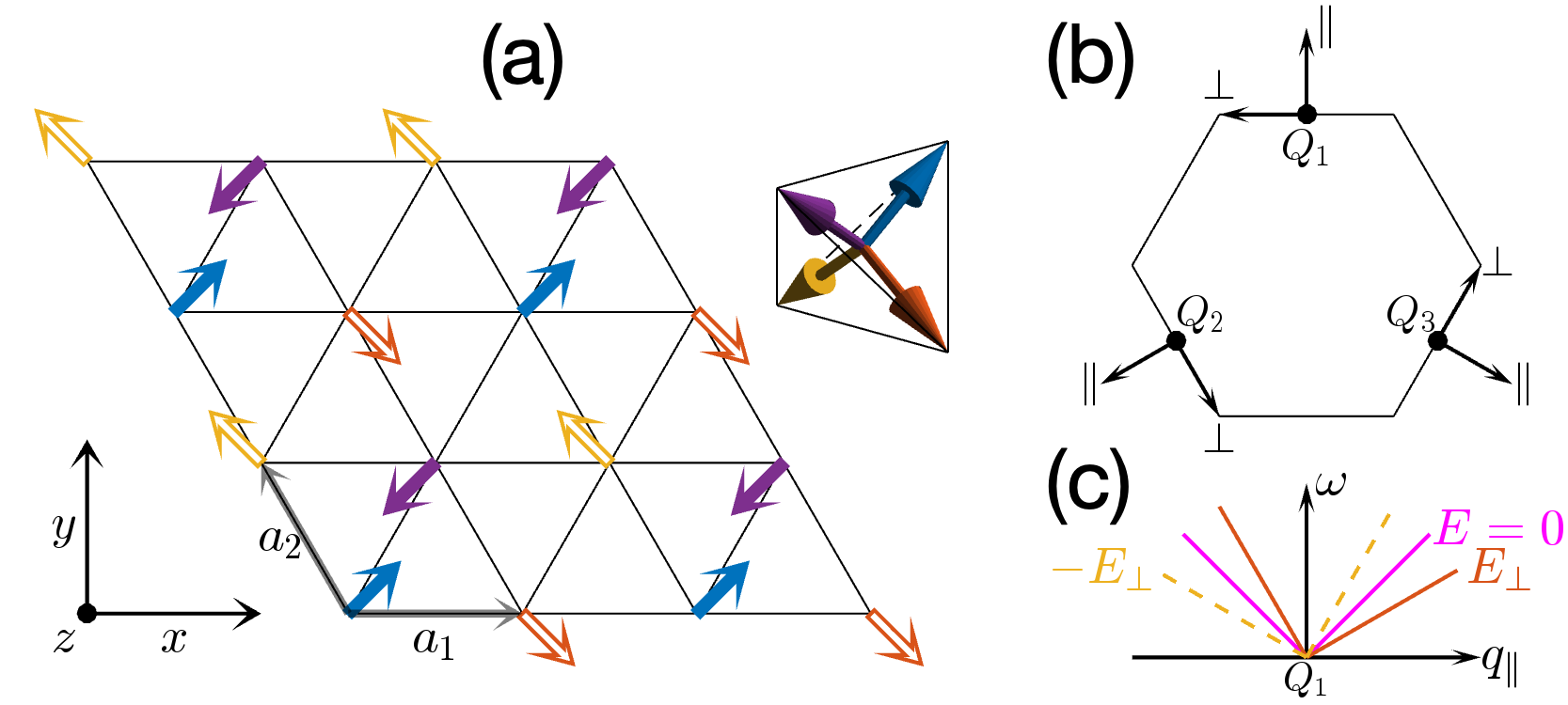}
\caption{(a) Top view of tetrahedral order in hexagonal lattice. Closed and open arrows denote spins above and below the paper plane. Distinct magnetic sublattices are colored differently. Insets show a side view of the spin vectors and the top view of spatial Cartesian coordinate system. (b) This order supports gapless spin waves at the M points ($Q_{1,2,3}$) of the lattice Brillouin zone. The longitudinal ($\parallel$) and transverse ($\perp$) directions are defined for each $Q$. (c) Spin wave dispersion relation near $Q_1$ (magenta solid line). It becomes non-reciprocal in an electric field in transverse direction (red solid line). Its non-reciprocity changes sign when the field is reversed (gold dashed line).}
\label{fig:sketch}
\end{figure}

Focusing on the zero SOC limit, we formulate symmetry constraints on electric field induced nonreciprocity. We require the spin wave spectra at wave vector $q$ and $-q$ are degenerate in zero field, and such degeneracy is lifted by electric field~\footnote{The spin waves may carry a good quantum number $\alpha$ in addition to wave vector, which can be used to label different modes. It may occur that the dispersion $\omega_\alpha(q) \neq \omega_\alpha(-q)$ for any given mode $\alpha$, but there exists a pair of modes $\alpha$ and $\alpha'$ such that $\omega_\alpha(q) = \omega_{\alpha'}(-q)$. This case does not fulfill the spectral non-reciprocity requirement and lies beyond the scope of this work.}. We find only a subset of \emph{noncoplanar} magnetic orders satisfy the constraints. A concrete example is the tetrahedral antiferromagnetic order in hexagonal lattice (Fig.\ref{fig:sketch}a)~\cite{Martin2008,Akagi2010,Kato2010,Kumar2010}. It supports three Goldstone spin wave modes at the M points (Fig.\ref{fig:sketch}b) of the first Brillouin zone (BZ). Effective field theory reveals that the Goldstone modes become nonreciprocal in electric field (Fig.\ref{fig:sketch}c) owing to a nonlinear magnetoelectric (ME) effect. We develop an adiabatic spin wave theory for magnetic insulators in weak electric field, which yields results that are fully consistent with the symmetry analysis and the field theory.

We examine the electric field induced nonreciprocity from the perspective of symmetry. Neglecting the subleading dipolar interactions, spin exchange interactions in zero SOC limit permit spin $SO(3)$ operations independent from the lattice ones. The spin space group (SSG) $G$ of a given magnetic order consists of combined lattice and spin operations that leaves the order invariant~\cite{Brinkman1966a,Brinkman1966b,Litvin1974,Smejkal2022,Corticelli2022,Liu2022,Jiang2024,Xiao2024}. Each element of $G$ is an ordered pair $(g_L, g_S)$, where $g_L$ and $g_S$ are respectively the lattice and spin operations. Note $g_S$ may be anti-unitary if it involves time reversal operations. For brevity, we refer to the spin space group element by its lattice part $g_L$ if no ambiguity occurs.

We show $G$ must fulfill stringent criteria to allow for electric field induced nonreciprocity. We consider two dimensional lattice for simplicity though our analysis can be extended to other dimensions. We assume that the electric field is along generic directions of the lattice, which lies in the $xy$ plane. 

First, the spectral reciprocity must be protected by an SSG element $g$ that is broken by electric field. This \emph{reciprocity criterion} ensures spin wave spectrum is reciprocal in the absence of electric field, but loses this property in electric field. 

Three types of symmetries can protect spectral reciprocity in zero field: (i) an anti-unitary onsite symmetry $(1,S)$, where $S$ is an anti-unitary spin operation; (ii) an anti-unitary translational symmetry $(T,S)$, where $T$ is a lattice translation and $S$ anti-unitary; (iii) a unitary 2-fold rotation about the $z$ axis $(C^z_2,R)$, where $R$ is unitary. We do not consider spatial inversion symmetry separately because the spatial inversion is identical to $C^z_2$ in two dimensions. All of these symmetries map the wave vector $q\to -q$ and thus protect reciprocity. 

The reciprocity criterion singlets out (iii) as the only compatible symmetry. The collinear and coplanar magnetic orders fall into the case (i)~\cite{Litvin1974}. For collinear orders, assuming the spins $\parallel \hat{e}^{\snum3}$ axis, $S = \Theta R(\hat{e}^{\snum1},\pi)$, where $\Theta$ is the time reversal operation, and $R(\hat{e}^{\snum1},\pi)$ is a $\pi$ rotation about $\hat{e}^{\snum1}$; for coplanar orders in the $\hat{e}^{\snum1}$-$\hat{e}^{\snum2}$ plane, $S = \Theta R(\hat{e}^{\snum3},\pi)$. However, as the electric field does not break these symmetries, it cannot induce nonreciprocity. By the same token, the reciprocity criterion rules out the case (ii). Finally, the case (iii) fulfills the reciprocity criterion in that the unitary $C^z_2$ rotation is broken by the electric field. 

The reciprocity criterion narrows down the candidate magnetic orders to noncoplanar ones with unitary translations and $C^z_2$ rotations. While it ensures field induced nonreciprocity, it is silent on the fate of the magnetic order in electric field. If the magnetic order loses its stability in an infinitesimal electric field, the spectral reciprocity, or the lack thereof, is no longer the inherent property of the zero field magnetic order but rather the new ground state. We therefore pose the \emph{stability criterion}: the magnetic order must be stable in the presence of an infinitesimal electric field.

We consider the Goldstone modes in order to apply this criterion. In a noncoplanar order, the Goldstone modes develop from position dependent rotations of the magnetic order, which we parametrize by a smooth function $\theta^\alpha(r)$. $\theta^\alpha$ is the axis-angle representation of the spin rotation~\cite{Halperin1977,Halperin1969}. $r$ is the real space position. $\theta^\alpha$ transforms as a vector under spin $SO(3)$ rotations and is even under time reversal. Furthermore, $\theta^\alpha$ carries finite momentum if SSG translations contain spin operation $g_S\neq 1$.

In zero field, the energy density due to such deformations is given by:
\begin{subequations}
\begin{align}
\mathcal{U} = \frac{1}{2}\rho^{\alpha\beta}_{jk}\partial_j\theta^\alpha \partial_k \theta^\beta.
\end{align}
Repeated indices are summed over. The superscripts $\alpha,\beta$ (subscripts $j,k$) label the Cartesian components in (spin) real space. The stiffness $\rho^{\alpha\beta}_{jk}$, viewed as a matrix with labels $j\alpha$ and $k\beta$, is real, symmetric, and positive definite. 

An infinitesimal electric field along a generic lattice direction break all SSG symmetries except the translations. As a result, new terms may appear in $\mathcal{U}$. Consider: 
\begin{align}
\eta^{\alpha\beta}_j (\theta^\alpha \partial_j \theta^\beta - \theta^\beta \partial_j \theta^\alpha),
\label{eq:linear_gradient}
\end{align}
\end{subequations}
where $\eta^{\alpha\beta}_j$ is a real coefficient that is skew-symmetric with respect to $\alpha,\beta$. Eq.\eqref{eq:linear_gradient} is the only first order spatial derivative term that can be added to $\mathcal{U}$. Adding this term results in an instability at $q\sim O(\eta/\rho)$. Terms with higher order spatial derivatives do not induce instability at infinitesimal coupling. 

Therefore, stability requires that Eq.~\eqref{eq:linear_gradient} is forbidden by translation symmetry. Consider the generators of the SSG translations $(T_1,R_1)$ and $(T_2,R_2)$. $R_1$ and $R_2$ are unitary as per the reciprocity criterion. As $T_1,T_2$ commute, so do $R_1$ and $R_2$~\cite{Messio2011}. Two possibilities arise:
\begin{enumerate}
\item $R_1$ and $R_2$ are coaxial rotations. We may set $R_1 = R(\hat{e}^{\snum3},\phi_1)$ and $R_2 = R(\hat{e}^{\snum3},\phi_2)$. $\theta^{\snum1}\pm i\theta^{\snum2}$ carry momentum $\pm (\frac{\phi_1}{2\pi}g_1+\frac{\phi_2}{2\pi}g_2)$, $g_{1,2}$ being primitive reciprocal lattice vectors. The term $\theta^{\snum1}\partial_i \theta^{\snum2} - \theta^{\snum2}\partial_i\theta^{\snum1}$ is allowed. 

\item $R_1$ and $R_2$ are $\pi$ rotations about two orthogonal axes. We set $R_1 = R(\hat{e}^{\snum1},\pi)$ and $R_2 = R(\hat{e}^{\snum2},\pi)$. $\theta^{\snum1}$, $\theta^{\snum2}$, and $\theta^{\snum3}$ carry momenta $g_2/2$, $g_1/2$, and $(g_1+g_2)/2$, respectively. Eq.~\eqref{eq:linear_gradient} is forbidden.
\end{enumerate}
Thus, the magnetic order is stable when the generators of SSG translations contain $\pi$ spin rotations about two orthogonal axes. Magnetic orders in the first category are unstable in electric field; examples include the 120$^{\circ}$ order in hexagonal lattice and its noncoplanar analogs~\cite{SM}.

To summarize, in two dimensions and without SOC, spin waves exhibits electric field induced nonreciprocity only if the magnetic order is noncoplanar, features a unitary $C^z_2$ symmetry, and the generators of its SSG translation symmetries contain $\pi$-rotations about two orthogonal spin axes. 

We find a set of  magnetic orders fulfill these requirements, including the tetrahedral order in hexagonal lattice (Fig.~\ref{fig:sketch}a). More examples shall be discussed later. Its spin configuration $S(r) \sim \hat{e}^{\snum1} e^{iQ_1\cdot r} + \hat{e}^{\snum2} e^{iQ_2 \cdot r} + \hat{e}^{\snum3} e^{iQ_3 \cdot r}$, where $r$ is the lattice position. $Q_{1,2,3}$ are the wave vectors at M points (Fig.\ref{fig:sketch}b). The SSG is generated by the following elements: unitary translations $(T(a_1),R(\hat{e}^{\snum1},\pi))$, $(T(a_2),R(\hat{e}^{\snum2},\pi))$, where $a_{1,2}$ are primitive lattice vectors; unitary $C^z_6$ rotation $(C^z_6,R([111],-\frac{2\pi}{3}))$; anti-unitary $C^x_2$ rotation $(C^x_2,\Theta R([0\overline{1}1],\pi))$. 

Having identified a candidate magnetic order, we are ready to investigate its spin waves. For now, we consider the Goldstone modes. Their dynamics is captured by an effective field theory whose dynamical variables are the spin rotations $\theta^\alpha(r)$ and the magnetization $m^\alpha(r)$~\cite{Halperin1977,Halperin1969}. They form canonical conjugate pairs. At quadratic level, the Lagrangian density assumes the general form, $\mathcal{L} = \omega^{\alpha\beta}m^\alpha \partial_t\theta^\beta - \mathcal{U}$. The first term encodes the Poisson bracket between $\theta^\alpha$ and $m^\alpha$. $\mathcal{U}$ is the potential energy, which is a quadratic form in $m^\alpha$ and $\partial_i\theta^\alpha$.

We constrain $\mathcal{L}$ by the SSG of the tetrahedral order. $\theta^\alpha$ and $m^\alpha$ transform non-trivially under the SSG. Both $\theta^\alpha$ and $m^\alpha$ carry the momentum $Q_\alpha$. It is convenient to define the longitudinal and transverse directions with respect to a given $Q_\alpha$ wave vector (Fig.\ref{fig:sketch}b). $(\theta^{\snum1},\theta^{\snum2},\theta^{\snum3}) \to -(\theta^{\snum1},\theta^{\snum3},\theta^{\snum2})$ under the anti-unitary $C^{x,y}_2$ rotations, whereas $(m^{\snum1},m^{\snum2},m^{\snum3}) \to (m^{\snum1},m^{\snum3},m^{\snum2})$ under these symmetries. The unitary $C^z_6$ rotations permute fields with different $\alpha$ label.

The Lagrangian density reads:
\begin{subequations}
\begin{align}
\mathcal{L} = m^\alpha \partial_t \theta^\alpha - \mathcal{U}_0 - \mathcal{U}_1.
\end{align}
The coefficient $\omega^{\alpha\beta} = \omega\delta^{\alpha\beta}$ because (a) translation symmetry forbids coupling of $m^\alpha \partial_t \theta^\beta$ when $\alpha\neq \beta$, and (b) the $C^z_6$ symmetry requires the coupling is the same for different $\alpha$. The constant $\omega$ is subsumed in $m^\alpha$.

$\mathcal{U}_0$ the potential energy in zero electric field. We deduce its form by similar arguments:
\begin{align}
\mathcal{U}_0 = \frac{\rho_{\spara}}{2} \partial_{\spara} \theta^\alpha \partial_{\spara} \theta^\alpha + \frac{\rho_{\sperp}}{2} \partial_{\sperp} \theta^\alpha \partial_{\sperp} \theta^\alpha + \frac{m^\alpha m^\alpha}{2\chi}.
\end{align}
The directional derivatives $\partial_{\spara},\partial_{\sperp}$ are defined with respect to a given $Q_\alpha$. Mixed derivative term $\partial_{\spara}\theta^\alpha \partial_{\sperp}\theta^\alpha$ is forbidden by $C^{x,y}_2$. $\rho_{\spara,\sperp}$ are the spin stiffness. $\chi$ is the magnetic susceptibility.

$\mathcal{U}_1$ is first order in electric field:
\begin{align}
\mathcal{U}_1 = -\zeta_1 E_{\sperp} m^\alpha \partial_{\spara} \theta^\alpha - \zeta_2 E_{\spara} m^\alpha \partial_{\sperp}\theta^\alpha.
\label{eq:m_partial_theta}
\end{align}
\label{eq:Lagrangian}
\end{subequations}
Eq.\eqref{eq:m_partial_theta} couples magnetization $m^\alpha$ to spin twist $\partial_i\theta^\alpha$, and will give rise to spin wave nonreciprocity. $E_{\spara,\sperp}$ are the projection of the electric field in the longitudinal and transverse directions with respect to $Q_\alpha$. $\zeta_{1,2}$ are free parameters. Eq.~\eqref{eq:m_partial_theta} may be viewed as a nonlinear ME effect where a combination of magnetization and twist of magnetic order produces electric polarization.

Following the standard procedure~\cite{SM}, we obtain from Eq.~\eqref{eq:Lagrangian} the spin wave dispersion relation:
\begin{align}
\omega^\alpha(q) = \zeta_1E_{\spara} q_{\sperp}+\zeta_2 E_{\sperp} q_{\spara} + \sqrt{\frac{\rho_{\spara} q^2_{\spara} + \rho_{\sperp} q^2_{\sperp}}{\chi}}.
\end{align}
Each $\theta^\alpha$ gives rise to an independent mode. We quantify the nonreciprocity by the dimensionless ratio:
\begin{align}
\frac{v_{\spara+} - v_{\spara-}}{v_{\spara+} + v_{\spara-}} = \zeta_2\sqrt{\frac{\chi}{\rho_{\spara}}}E_{\sperp};
\,
\frac{v_{\sperp+} - v_{\sperp-}}{v_{\sperp+} + v_{\sperp-}} = \zeta_1\sqrt{\frac{\chi}{\rho_{\sperp}}}E_{\spara}.
\label{eq:v_ratio}
\end{align}
$v_{\spara\pm}$ are respectively the group velocity along (+) and against (-) the longitudinal direction. $v_{\sperp\pm}$ are similarly defined. The direction and magnitude of nonreciprocity in the longitudinal direction is controlled by the electric field in the transverse direction (Fig.~\ref{fig:sketch}c), and vice versa. 

\begin{figure}
\centering
\includegraphics[width  = \columnwidth]{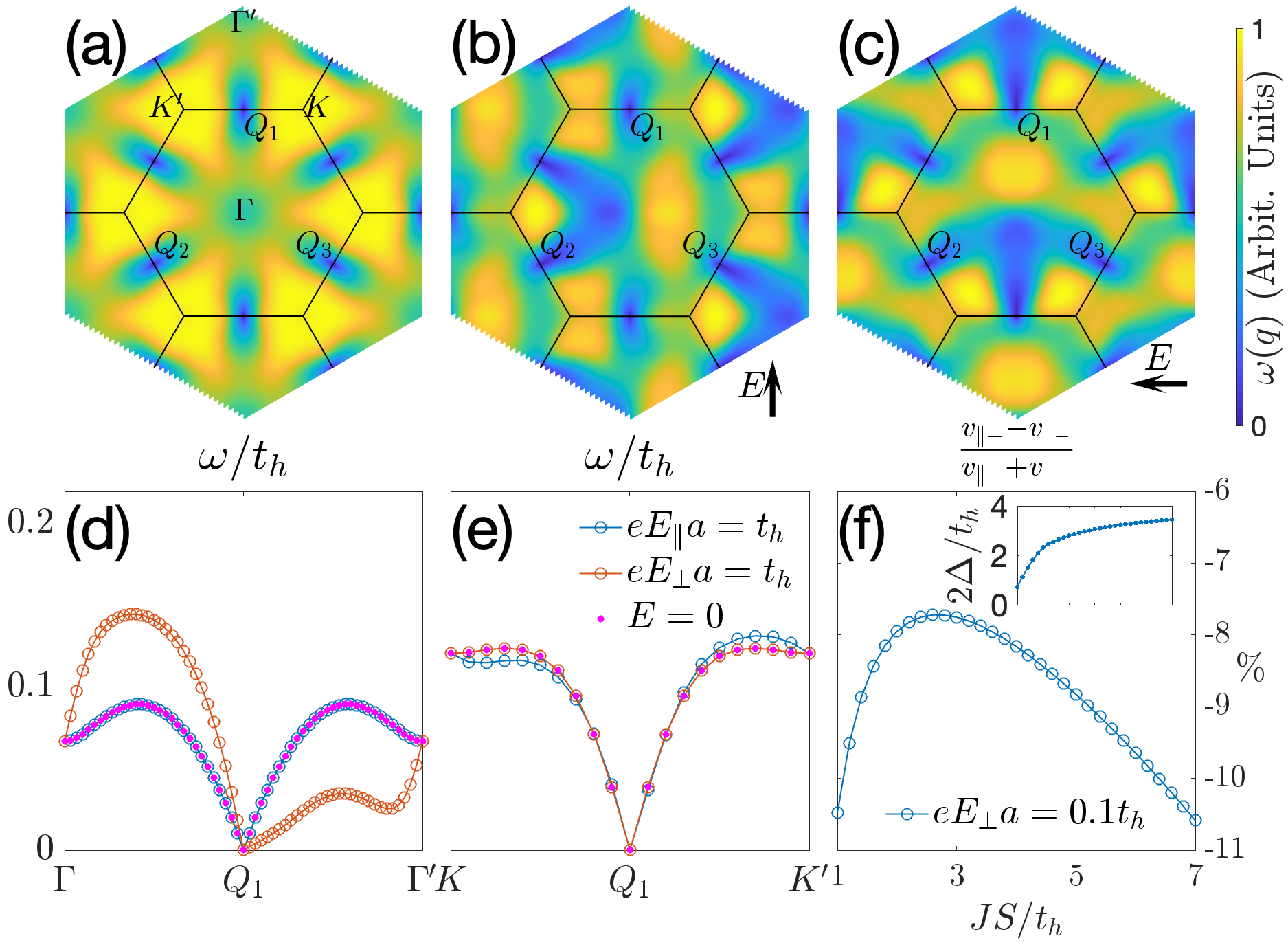}
\caption{(a) Spin wave dispersion $\omega(q)$ in the tetrahedral order hosted by the classical Kondo model (Eq.~\eqref{eq:kondo_model}) in hexagonal lattice at $J/t_h = 2$ and $1/4$ electron filling. The lattice Brillouin zone boundary and selected high symmetry points are marked. The $\omega(q)$ maximum is scaled to 1. (b) $\omega(q)$ in an electric field $eE_xa/t_h = 1$. (c) $\omega(q)$ with $eE_ya/t_h=-1$. (d) $\omega(q)$ along the longitudinal direction of $Q_1$. (e) Similar to (d) but for the transverse direction. (f) Ratio of group velocities at $Q_1$ along the longitudinal direction plotted as a function of $J/t_h$. Electric field $eE_\perp a/t_h = 0.1$. Inset: the electron particle-hole gap $2\Delta$ as a function of $J/t_h$.}
\label{fig:disp}
\end{figure}

We corroborate the field theory results by a microscopic model calculation. Consider the classical Kondo model on hexagonal lattice~\cite{Martin2008}:
\begin{align}
H = -t_{rr'} c^\dagger_{rs} c^{\phantom\dagger}_{r's} -JS \hat{n}^\alpha_r  \sigma^\alpha_{ss'} c^\dagger_{rs} c^{\phantom\dagger}_{rs'}.
\label{eq:kondo_model}
\end{align} 
Repeated indices are summed over. $c_{rs}$ ($c^\dagger_{rs}$) annihilates (creates) an electron on lattice site $r$ with spin state $s=\uparrow,\downarrow$. The hopping integral $t_{rr'} = t_h>0$ if $r,r'$ are nearest neighbors, and $t_{rr'} = 0$ otherwise. The exchange constant $J>0$. $S$ is the spin length. $\hat{n}_r$ is the directional vector of the classical spin on site $r$.  $\sigma^\alpha$ are the Pauli matrices. The spins obey the Landau-Lifshitz equation.

Eq.~\eqref{eq:kondo_model} exhibits a rich ground state phase diagram as a function of filling factor $\nu$ and parameter ratio $J/t_h$~\cite{Akagi2010}. It is sufficient for our purpose to focus on $\nu = 1/4$~\cite{Kato2010}. For a wide range of $J/t_h$, the ground state is an insulator with tetrahedral order. We set $J/t_h=2$ and $S=1$ unless stated otherwise.

We employ the adiabatic spin wave theory~\cite{Niu1998,Halilov1998,Zhou2025} to treat the spin dynamics in electric field. In the velocity gauge, a uniform electric field is described by the Peierls substitution $t_{rr'} \to t_{rr'}\exp[i eA_j(t)(r_j-r'_j)]$, where $A_j(t)$ is the vector potential. The electric field $E_j = -\partial_t A_j$. $e=-|e|$ is the electron charge. 

The following Lagrangian captures the dynamics of the system:
\begin{align}
L = \mathcal{A}^\alpha_r \partial_t \hat{n}^\alpha_r + i\langle \Psi| \partial_t\Psi\rangle - \langle\Psi|H|\Psi\rangle.
\end{align} 
The Berry connection $\mathcal{A}^\alpha_r = \mathcal{A}^\alpha(\hat{n}_r)$ is a function of $\hat{n}_r$, satisfying $\nabla_{\hat{n}}\times \mathcal{A} = -S\hat{n}$. $|\Psi\rangle$ is the electron state. $H$ depends on classical spin configuration $\{\hat{n}\}$ and the vector potential $A(t)$. 

Provided the electric field ($eEa$, where $a$ is lattice spacing) and the spin wave frequency $\omega(q)$ are much less than the particle-hole excitation gap, $\{\hat{n}\}$ and $A(t)$ are adiabatic perturbations to $H$. We thus postulate $|\Psi\rangle = |\Psi_G\rangle$, where $|\Psi_G\rangle$ is the instantaneous ground state of $H$ for given $\{\hat{n}\}$ and $A$. Substituting this ans\"{a}tz in, we obtain the effective spin Lagrangian:
\begin{align}
L = \mathcal{A}'^{\alpha}_r \partial_t \hat{n}^\alpha_r -\mathcal{E}_G + E_j P_j.
\label{eq:L_spin}
\end{align}
$\mathcal{A}'^{\alpha}_r = \mathcal{A}(\hat{n}_r) + i\langle\Psi_G|\nabla^\alpha_{\hat{n}_r}\Psi_G\rangle$ accounts for the electron contribution to the spin Berry connection. $\mathcal{E}_G$ is the instantaneous ground state energy. $P_j = -i\langle \Psi_G|\partial_{A_j}\Psi_G\rangle$ is the electron polarization~\cite{Resta2007,Coh2009}~\footnote{The definition of electric polarization for a Chern insulator is subtle~\cite{Coh2009}. This subtlety does not impact the problem at hand as it concerns the change in polarization due to spatially smooth deformation of $\{\hat{n}\}$.}. 

We linearize Eq.~\eqref{eq:L_spin} about the ground state. We write $\hat{n}_r = \hat{e}^{\snum1}_r u^{\snum1}_r + \hat{e}^{\snum2}_r u^{\snum2}_r + \hat{e}^{\snum3}_r (1-u^2/2)$. $\hat{e}^{\alpha}_r$ form the local spin frame on site $r$. $\hat{e}^{\snum3}_r$ coincides with $\hat{n}_r$ in the ground state. $u^\alpha_r\ll 1$ parametrizes the deviation from the ground state. Expanding Eq.~\eqref{eq:L_spin} to quadratic order in $u^\alpha_r$ and first order in $A_j$ yields~\cite{SM}:
\begin{subequations}
\begin{align}
L = \frac{1}{2}  (\Omega^{\alpha\beta}_{rr'} u^\alpha_r \partial_t u^\beta_{r'} - \Phi^{\alpha\beta}_{rr'} u^\alpha_r u^\beta_{r'} + E_jZ^{\alpha\beta}_{jrr'} u^\alpha_r u^\beta_{r'}).
\end{align}
The Berry curvature is given by,
\begin{align}
\Omega^{\alpha\beta}_{rr'} = -S\delta_{rr'}\epsilon^{\alpha\beta} + F[u^\alpha_r,u^\beta_{r'}].
\end{align}
$\epsilon^{\alpha\beta}$ is the Levi-Civita symbol. The notation $F[X,Y] = i(\langle \nabla_X \Psi_G|\nabla_Y \Psi_G\rangle - \langle \nabla_Y \Psi_G|\nabla_X \Psi_G\rangle)$ stands for the electron Berry curvature with respect to parameters $X$ and $Y$. The Hessian reads,
\begin{align}
\Phi^{\alpha\beta}_{rr'} = \frac{\partial^2 \mathcal{E}_G}{ \partial u^{\alpha}_r \partial u^{\beta}_{r'}}.
\end{align}
$Z^{\alpha\beta}_{jrr'}$ is a nonlinear ME coefficient:
\begin{align}
Z^{\alpha\beta}_{jrr'} = \frac{1}{2}(\frac{\partial F[A_j,u^\alpha_r ]}{\partial u^\beta_{r'}}+\frac{\partial F[A_j,u^\beta_{r'}]}{\partial u^\alpha_r}),
\end{align}
\label{eq:L_linear}
\end{subequations}
which plays a role similar to $\zeta_{1,2}$ in Eq.~\eqref{eq:Lagrangian}. $\Omega,\Phi,Z$ are calculated numerically from the electron Bloch states at equilibrium. The spin wave dispersion is found by solving the spin wave equation deduced from Eq.~\eqref{eq:L_linear}~\cite{SM}. 

Fig.~\ref{fig:disp}a shows the spin wave dispersion $\omega(q)$ in the \emph{lattice} BZ in zero field. Gapless at the M points, $\omega(q)$ is six-fold symmetric, reflecting the high symmetry of the tetrahedral order. Applying an electric field in $y$ results in a nonreciprocal dispersion (Fig.~\ref{fig:disp}b). It retains the reflection symmetry about $x$ due to an anti-unitary $C^y_2$ that maps $(q^x,q^y)\to (q^x,-q^y)$. Likewise, an electric field in $-x$ has similar effect (Fig.~\ref{fig:disp}c). $\omega(q)$ is reflection symmetric about $y$ due to the anti-unitary $C^x_2$ symmetry. 

Focusing on the Goldstone mode near $Q_1$, $\omega(q)$ is reciprocal in zero field along the longitudinal (Fig.~\ref{fig:disp}d) direction. A transverse field induces nonreciprocity in the said direction, but a longitudinal field does not. $\omega(q)$ in the transverse direction exhibits similar behavior, with the role of transverse/longitudinal exchanged, and to a much lesser degree (Fig.~\ref{fig:disp}e).

We quantify the nonreciprocity by the dimensionless velocity ratio (Eq.~\eqref{eq:v_ratio}). Fig.~\ref{fig:disp}f shows the ratio in the longitudinal direction at $Q_1$ in $eE_\perp a = 0.1t_h$. It exhibits a modest dependence on $J/t_h$. Both the electric field and the spin wave bandwidth is below the electron particle-hole excitation gap $2\Delta$ (Fig.~\ref{fig:disp}f, inset), justifying the use of adiabatic spin wave theory.

\begin{figure}
\centering
\includegraphics[width  = \columnwidth]{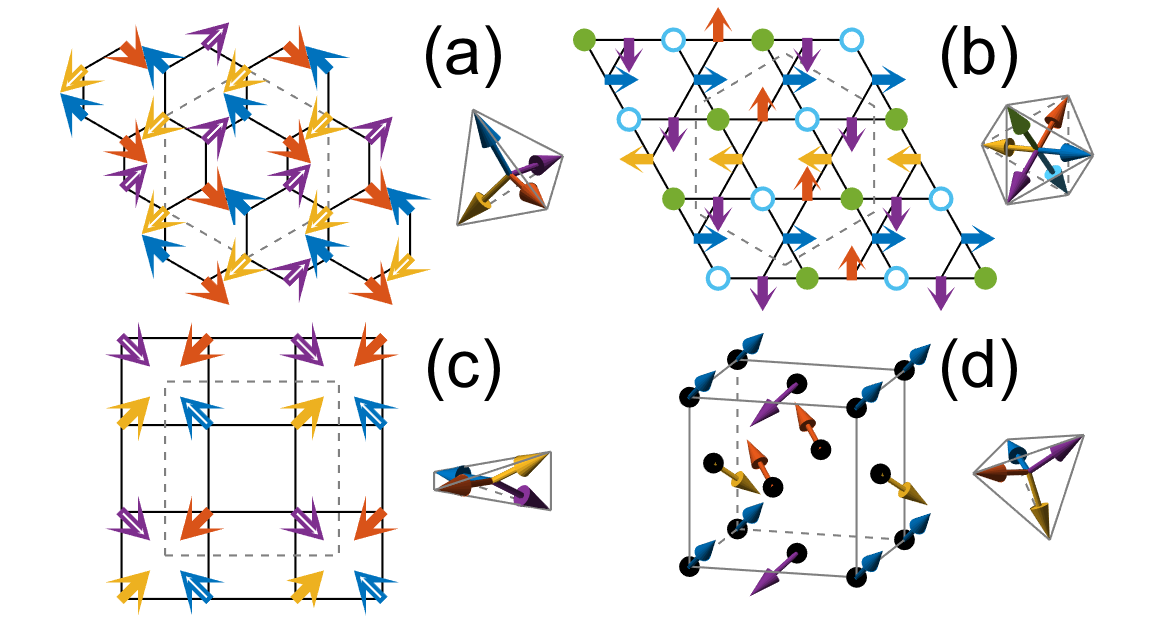}
\caption{Further examples of non-coplanar magnetic orders exhibiting electric-field induced non-reciprocity. (a) Tetrahedral order in honeycomb. (b) Octahedral order in kagome. (c) AF umbrella order in square lattice. (d) Tetrahedral order in face-centered cubic lattice. Nomenclature of a$\sim$c follows Ref.~\onlinecite{Messio2011}. Dashed lines mark magnetic unit cell.}
\label{fig:examples}
\end{figure}

The microscopic model calculation thus confirms that the spin waves in tetrahedral order in hexagonal lattice exhibits non-reciprocity in an electric field. The results are fully consistent with the symmetry analysis and the field theory. Using the symmetry conditions found in the beginning of this work, we have found more examples (Fig.~\ref{fig:examples}(a$\sim$c)) by going through~\cite{SM} the regular magnetic orders tabulated in Ref.~\onlinecite{Messio2011}. 

Our symmetry analysis may be straightforwardly generalized to three dimensions. We rule out the collinear and coplanar orders and anti-unitary SSG translations. The unitary $C^z_2$ symmetry is now replaced by unitary inversion. For stability, the translation generators must be accompanied by mutually orthogonal $\pi$-rotations in spin space. A concrete example is the tetrahedral order in face-centered cubic (FCC) lattice (Fig.~\ref{fig:examples}d).

These magnetic orders are predicted to exist in numerous systems including doped graphene~\cite{Makogon2011,Li2012,Kiesel2012,Wang2012,Jiang2014,Jiang2015}, Mori\'{e} superlattices~\cite{Liu2018,He2023,Wilhelm2023}, and van der Waals magnets~\cite{Jiang2025}. The tetrahedral order is observed in hexagonal GdGaI~\cite{Okuma2024} and FCC MnTe$_2$~\cite{Burlet1997,Zhu2024}. The nonreciprocal spin waves with $qa\ll1$ may be probed in a small, strip/rod shaped sample electrically~\cite{Devolder2021,Kanj2023} or optically~\cite{Hashimoto2017}. An electric field across the strip/rod results in nonreciprocal dispersion in the length direction. As a crude estimate, assuming $t_h\sim 1$eV and $a\sim 5$\AA, an electric field $\sim0.2$V/nm can induce $\sim$10\% velocity difference (Fig.~\ref{fig:disp}f). This field is strong but within the reach of the present techniques~\cite{Weintrub2022}. Systems with small bandwidth and large $a$, such as Moir\'{e} superlattices, may require much weaker electric field.

Our analysis focuses on the zero SOC limit. A weak SOC $\lambda\ll1$ breaks the spin $SO(3)$ symmetry.  The nonreciprocity criterion is nevertheless robust since it only concerns the time reversal part of the spin operation. However, the on-site anti-unitary symmetries unique to collinear/coplanar magnetic orders are lost.  Meanwhile, the Goldstone modes acquire a gap. The magnetic orders that violate the stability criterion are stable up to a small critical field $E_c$, which vanishes as $\lambda\to 0$. 

To conclude, we have shown that the electric field can generate and control nonreciprocal spin waves in noncoplanar magnetic insulators. The electric field operates through a nonlinear ME effect instead of SOC. Our work opens up a few research directions. First, the symmetry analysis can be carried out for more specialized cases such as $E\parallel$ high symmetry directions. Second, the adiabatic spin wave theory, integrated with the density functional theory, permits a systematic search for ideal candidate materials. Third, the nonlinear ME effect may be mediated by phonons or electron collective modes, which could enhance the nonreciprocity. Finally, our work demonstrates the strength of SSG, combined with field theory, in constraining the Goldstone mode dynamics. It will be fruitful to extend the present analysis to hydrodynamics~\cite{Halperin1977,Halperin1969} to account for dissipative effects.

\begin{acknowledgments}
This work is supported by the National Key Research and Development Program of China (Grants No. 2021YFA1403800, No. 2024YFA1408700), the National Natural Science Foundation of China (Grant No. 12250008), and the National Science Foundation through award No. NSF DMR-2011876. This work was initiated at Aspen Center for Physics, which is supported by National Science Foundation grant PHY-2210452. We thank the hospitality of the Institute for Solid State Physics at University of Tokyo, where a part of the work was carried out.
\end{acknowledgments}

\bibliography{tripleQ_electric.bib}

\appendix
\onecolumngrid
\clearpage

\makeatletter
\renewcommand{\c@secnumdepth}{0}
\makeatother

\numberwithin{figure}{section}

\renewcommand{\thefigure}{\Alph{section}.\arabic{figure}}
\renewcommand{\thetable}{\Alph{section}.\arabic{table}}
\renewcommand{\theequation}{\Alph{section}.\arabic{equation}}

\section{Spin space group of noncoplanar regular magnetic orders}

\begin{table}
\centering
\begin{tabular}{cccccc}
\hline
Lattice & RMO & $T_1$ & $T_2$ & $C^z_2$ & Nonreciprocity \\
\hline
Hexagonal & Tetrahedral & $R(\hat{e}^{\snum1},\pi)$ & $R(\hat{e}^{\snum2},\pi)$  & $1$ & $\checkmark$ \\
- & F-umbrella & $R(\hat{e}^{\snum3 },-\frac{2\pi}{3})$ & $R(\hat{e}^{\snum 3},-\frac{2\pi}{3})$ & $\Theta R(\hat{e}^{\snum2},\pi)$ & NR in zero field/Unstable  \\
Honeycomb & Cubic & $R(\hat{e}^{\snum1},\pi)$ & $R(\hat{e}^{\snum2},\pi)$  & $\Theta$ & NR in zero field  \\
- & Tetrahedral & $R(\hat{e}^{\snum1},\pi)$ & $R(\hat{e}^{\snum2},\pi)$  & $1$ & $\checkmark$ \\
Kagome & Octahedral & $R(\hat{e}^{\snum1},\pi)$ & $R(\hat{e}^{\snum2},\pi)$  & $1$ & $\checkmark$ \\
- & Cuboc1  & $R(\hat{e}^{\snum1},\pi)$ & $R(\hat{e}^{\snum2},\pi)$ & $\Theta$ & NR in zero field \\
- & Cuboc2 & $R(\hat{e}^{\snum1},\pi)$ & $R(\hat{e}^{\snum2},\pi)$ & $\Theta$ & NR in zero field \\
- & $q=0$ Umbrella & $1$ & $1$ & $1$  & Unstable \\
- & $\sqrt{3}\times \sqrt{3}$ Umbrella & $R(\hat{e}^{\snum3},-\frac{2\pi}{3})$ & $R(\hat{e}^{\snum3},-\frac{2\pi}{3})$ & $\Theta R(\frac{\sqrt{3}}{2}\hat{e}^{\snum1}-\frac{1}{2}\hat{e}^{\snum2},\pi)$ & NR in zero field/Unstable \\
Square & Tetrahedral umbrella & $R(\hat{e}^{\snum2},\pi)$ & $R(\hat{e}^{\snum1},\pi)$ & $1$  & $\checkmark$ \\
- & Umbrella  & $\Theta R(\hat{e}^{\snum1},\pi)$ & $\Theta R(\hat{e}^{\snum2},\pi)$ & $1$ & R in E field \\
\hline 
\end{tabular}
\caption{Spin space group (SSG) translations and $C^z_2$ rotations of noncoplanar regular magnetic orders (RMOs) in selected lattices. The last column lists their properties in relation to the spin wave nonreciprocity: unstable in infinitesimal electric field (``unstable"), nonreciprocal spin wave in zero electric field (``NR in zero field"), reciprocal spin wave in electric field (``R in E field"), and electric field-induced nonreciprocity (checkmark).}
\label{tab:SSG_RMO}
\end{table}

The regular magnetic orders (RMOs) are defined in Ref.~\cite{Messio2011} as magnetic orders that preserve the lattice symmetries modulo global spin transformations. Recall that the spin space group (SSG) $G$ is formed by symmetry operations of the form $(g_L,g_S)$, where $g_L$ is a lattice operation and $g_S$ a spin space operation. On one hand, all distinct space operations $g_L$ that appear in the SSG form a group $G_L$. On the other hand, the elements of the form $(1,g_S)$ form a normal subgroup $S$ of the SSG. It is easy to see that $G_L \cong G/S $. The definition for RMO is equivalent to the statement that $G/S$ is isomorphic to the space group of the lattice that hosts the said order. They constitute the most symmetric magnetic orders for a given lattice. 

In this section, we discuss the SSG of the noncoplanar regular magnetic order tabulated in Ref.~\cite{Messio2011} in relation to the spin-wave reciprocity. According to the discussion in the main text, the key symmetry elements are the SSG translations and the $C^z_2$ rotations. We list the accompanying spin space operations $g_S$ in Table~\ref{tab:SSG_RMO}. The notation is the same as the main text. $\Theta$ refers to the anti-unitary time reversal operation. $R(\hat{n},\theta)$ refers to a proper spin rotation about the $\hat{n}$ axis by an angle $\theta$.

We go through the RMOs one by one.

\subsection{Hexagonal lattice \label{sec:hexagonal}}

\begin{figure}
\centering
\includegraphics[width = \textwidth]{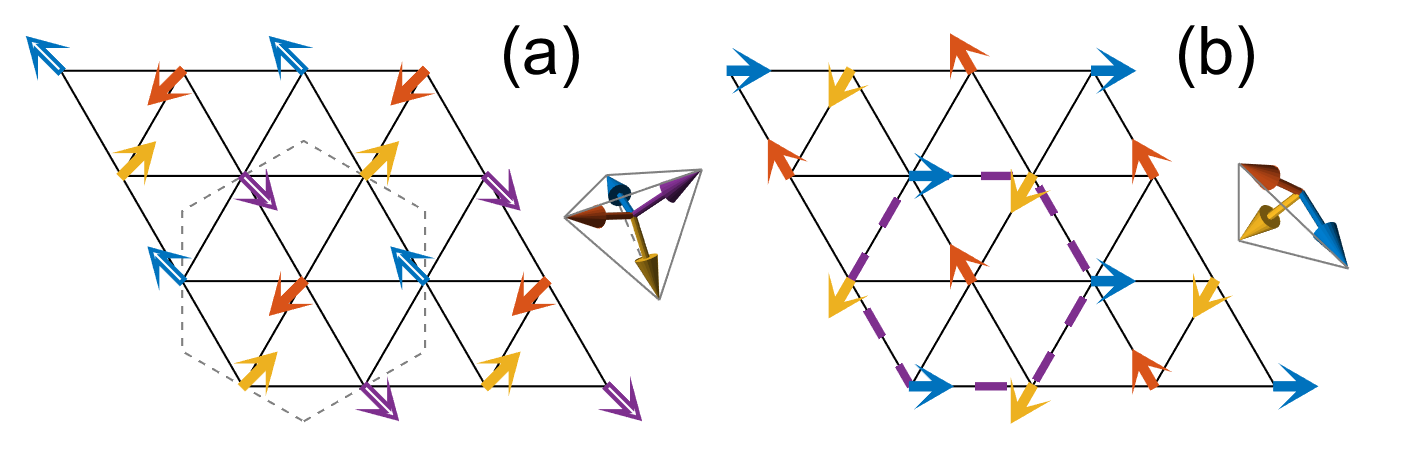}
\caption{(a) Top view of the hexagonal lattice tetrahedral order. Arrows indicate the direction of the spin. Closed and open arrows denote the spins above and below paper plane. The inset shows the side view of spin orientation. Gray dashed line marks the magnetic unit cell. (b) Hexagonal F-umbrella order. Thick purple dashed line marks the magnetic unit cell.}
\label{fig:RMO_hexagonal}
\end{figure}

We use the following convention for hexagonal Bravais lattice. It is generated by primitive lattice vectors: $a_1 = (1,0)$,$a_2 = (-\frac{1}{2},\frac{\sqrt{3}}{2})$. $T_1$ and $T_2$ refer to the lattice translations by $a_1$ and $a_2$, respectively. 
$\hat{e}^{\snum{1},\snum{2},\snum{3}}$ form a right-hand frame in the spin space. 

\begin{enumerate}
\item Hexagonal lattice tetrahedral order (Fig.~\ref{fig:RMO_hexagonal}a). Without loss of generality, we set the spin configuration to:
\begin{align}
S(r) = N(\hat{e}^{\snum1} e^{iQ_1\cdot r} + \hat{e}^{\snum2} e^{iQ_2\cdot r} + \hat{e}^{\snum3} e^{iQ_3\cdot r}).
\end{align}
$Q_{1,2,3}$ are the three wave vectors at the M points of the first Brillouin zone: $Q_1 = \frac{2\pi}{\sqrt{3}}(0,1)$, $Q_2 = \frac{2\pi}{\sqrt{3}}(-\frac{\sqrt{3}}{2},-\frac{1}{2})$, $Q_3 = \frac{2\pi}{\sqrt{3}}(\frac{\sqrt{3}}{2},-\frac{1}{2})$. $N$ is a constant. Under the lattice translation $T_1$, the spin configuration becomes:
\begin{align}
S'(r) = N(\hat{e}^{\snum1} e^{iQ_1\cdot r} - \hat{e}^{\snum2} e^{iQ_2\cdot r} - \hat{e}^{\snum3} e^{iQ_3\cdot r}).
\end{align}
$R(\hat{e}^{\snum1},\pi)$ restores $S'(r)\to S(r)$. We thus find the SSG translation generator $(T_1,R(\hat{e}^{\snum1},\pi))$. Likewise, the other SSG translation is $(T_2,R(\hat{e}^{\snum2},\pi))$. Finally, $S(r)$ is invariant under the lattice $C^z_2$ rotation about the origin: 
\begin{align}
S'(r) &= N(\hat{e}^{\snum1} e^{-iQ_1\cdot r} + \hat{e}^{\snum2} e^{-iQ_2\cdot r} + \hat{e}^{\snum3} e^{-iQ_3\cdot r}) 
\nonumber\\
&=  N(\hat{e}^{\snum1} e^{iQ_1\cdot r} + \hat{e}^{\snum2} e^{iQ_2\cdot r} + \hat{e}^{\snum3} e^{iQ_3\cdot r}) = S(r).
\end{align}
In the second line, we have used the fact that $\exp(iQ_{1,2,3}\cdot r) = \exp(-iQ_{1,2,3}\cdot r)$. This operation corresponding to the SSG element $(C^z_2,1)$.

\item Hexagonal lattice F-umbrella order (Fig.~\ref{fig:RMO_hexagonal}b). The spin configuration is given by:
\begin{align}
S(r) = N(\hat{e}^{\snum1} \cos(Q \cdot r) + \hat{e}^{\snum2} \sin(Q\cdot r)) + M \hat{e}^{\snum3} = N(\frac{\hat{e}^{\snum1}-i\hat{e}^{\snum2}}{2} e^{iQ \cdot r} + c.c.) + M \hat{e}^{\snum3}.
\end{align}
Here, $Q$ is at the K point of the Brillouin zone: $Q = \frac{4\pi}{3}(1,0)$. $N,M$ are constants. Acting the lattice translation $T_1$ on $S(r)$ transforms it to:
\begin{align}
S'(r) = N(\frac{\hat{e}^{\snum1}-i\hat{e}^{\snum2}}{2}e^{i\frac{2\pi}{3}} e^{iQ \cdot r} + c.c.) + M \hat{e}^{\snum3}.
\end{align}
The $-2\pi/3$ spin rotation about $\hat{e}^{\snum3}$ axis restores the magnetic order, yielding a SSG translation generator $(T_1,R(\hat{e}^{\snum3},-\frac{2\pi}{3}))$. By the same token, we can find the other generator $(T_2,R(\hat{e}^{\snum3},-\frac{2\pi}{3}))$. Acting lattice $C^z_2$ rotation on $S(r)$ maps it to:
\begin{align}
S'(r) = N(\hat{e}^{\snum1} \cos(Q \cdot r) - \hat{e}^{\snum2} \sin(Q\cdot r)) + M \hat{e}^{\snum3}. 
\end{align}
The spin operation $\Theta R(\hat{e}^{\snum2},\pi)$ restores the magnetic order. Thus, $(C^z_2,\Theta R(\hat{e}^{\snum2},\pi))$ is in the SSG.

\end{enumerate}

\subsection{Honeycomb lattice}

\begin{figure}
\centering
\includegraphics[width = \textwidth]{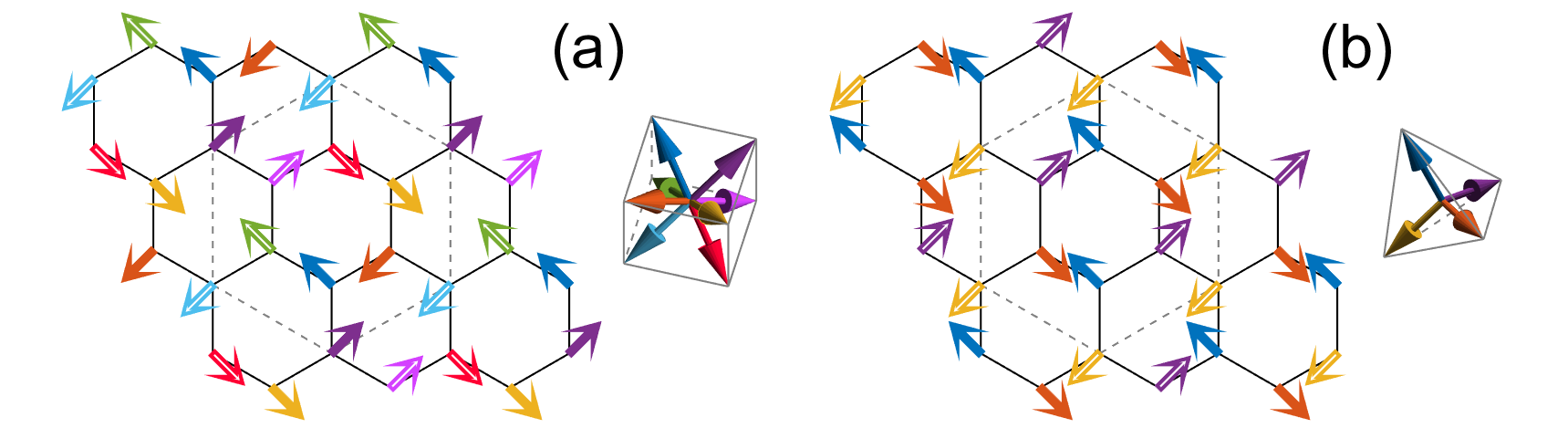}
\caption{(a) Honeycomb cubic order. (b) Honeycomb tetrahedral order. The convention is the same as Fig.~\ref{fig:RMO_hexagonal}.}
\label{fig:RMO_honeycomb}
\end{figure}

We use the following convention for the honeycomb lattice. The primitive lattice vectors: $a_1 = (1,0)$,$a_2 = (-\frac{1}{2},\frac{\sqrt{3}}{2})$. The origin is at the center of a hexagon. $b$ is a vector pointing from the origin to a neighboring A site: $b = \frac{1}{\sqrt{3}}(0,1)$. 

\begin{enumerate}
\item Honeycomb cubic order (Fig.~\ref{fig:RMO_honeycomb}a). The spin configuration is given by: 
\begin{align}
S(r) = \left\{ \begin{array}{cc}
N(\hat{e}^{\snum1} e^{iQ_1\cdot (r-b)} + \hat{e}^{\snum2} e^{iQ_2\cdot (r-b)} + \hat{e}^{\snum3} e^{iQ_3\cdot (r-b)}) & (r\in A) \\
-N(\hat{e}^{\snum1} e^{iQ_1\cdot (r+b)} + \hat{e}^{\snum2} e^{iQ_2\cdot (r+b)} + \hat{e}^{\snum3} e^{iQ_3\cdot (r+b)}) & (r\in B)
\end{array}\right. .
\end{align}
$Q_{1,2,3}$ are the wave vectors associated with the M points of the first Brillouin zone. Each sublattice hosts a hexagonal tetrahedral order. Since translation maps a sublattice to itself, the SSG translations are of the same form as the hexagonal tetrahedral order. The lattice $C^z_2$ rotation about the origin maps $r\to -r$ and exchanges A and B sublattices. As a result, the spin configuration becomes:
\begin{align}
S'(r) &= \left\{ \begin{array}{cc}
-N(\hat{e}^{\snum1} e^{iQ_1\cdot (-r+b)} + \hat{e}^{\snum2} e^{iQ_2\cdot (-r+b)} + \hat{e}^{\snum3} e^{iQ_3\cdot (-r+b)}) & (r\in A) \\
N(\hat{e}^{\snum1} e^{iQ_1\cdot (-r-b)} + \hat{e}^{\snum2} e^{iQ_2\cdot (-r-b)} + \hat{e}^{\snum3} e^{iQ_3\cdot (-r-b)}) & (r\in B)
\end{array}\right. 
\nonumber\\
& = \left\{ \begin{array}{cc}
-N(\hat{e}^{\snum1} e^{iQ_1\cdot (r-b)} + \hat{e}^{\snum2} e^{iQ_2\cdot (r-b)} + \hat{e}^{\snum3} e^{iQ_3\cdot (r-b)}) & (r\in A) \\
N(\hat{e}^{\snum1} e^{iQ_1\cdot (r+b)} + \hat{e}^{\snum2} e^{iQ_2\cdot (r+b)} + \hat{e}^{\snum3} e^{iQ_3\cdot (r+b)}) & (r\in B)
\end{array}\right. .
\end{align}
In the second line, we have used $\exp(iQ_{1,2,3}\cdot (r-b)) = \exp(-iQ_{1,2,3}\cdot (r-b))$ for $r\in A$, and $\exp(iQ_{1,2,3}\cdot (r+b)) = \exp(-iQ_{1,2,3}\cdot (r+b))$ for $r\in B$. The time reversal $\Theta$ restores the magnetic order. Thus, the SSG contains the element $(C^z_2,\Theta)$.

\item Honeycomb tetrahedral order (Fig.~\ref{fig:RMO_honeycomb}b). The spin configuration is given by: 
\begin{align}
S(r) = \left\{ \begin{array}{cc}
N(\hat{e}^{\snum1} e^{iQ_1\cdot (r-b)} + \hat{e}^{\snum2} e^{iQ_2\cdot (r-b)} + \hat{e}^{\snum3} e^{iQ_3\cdot (r-b)}) & (r\in A) \\
N(\hat{e}^{\snum1} e^{iQ_1\cdot (r+b)} + \hat{e}^{\snum2} e^{iQ_2\cdot (r+b)} + \hat{e}^{\snum3} e^{iQ_3\cdot (r+b)}) & (r\in B)
\end{array}\right. .
\end{align}
This order is very similar to that of the honeycomb cubic order except that its symmetric with respect to exchanging sublattices A and B. Thus, its SSG translations are identical to that of the honeycomb cubic order, and its SSG $C^z_2$ rotation is given by $(C^z_2,1)$.
\end{enumerate}

\subsection{Kagome lattice}

\begin{figure}
\centering
\includegraphics[width = 0.8\textwidth]{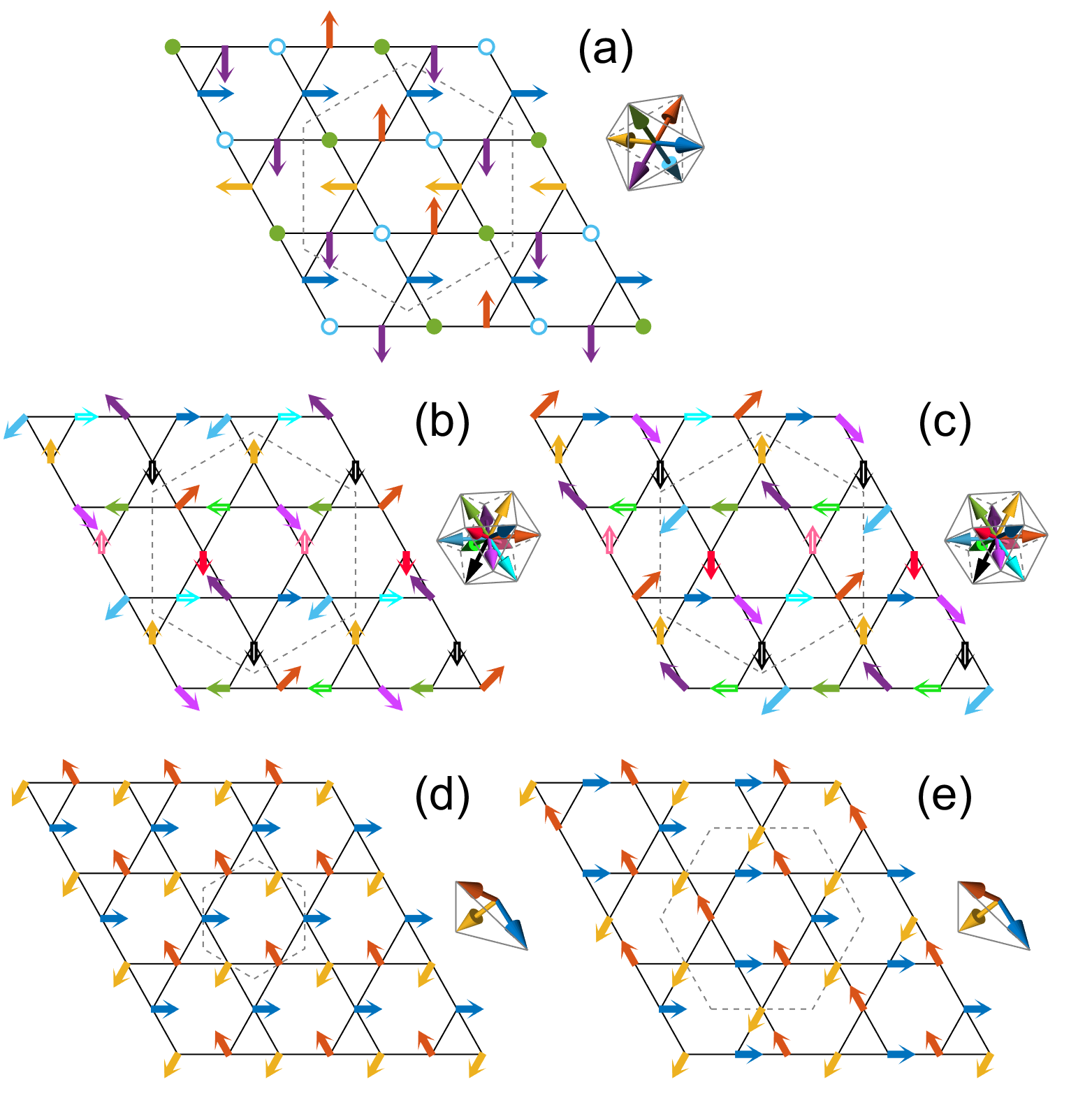}
\caption{(a) Kagome octahedral order. (b) Kagome cuboc1 order. (c) Kagome cuboc2 order. (d) Kagome $q=0$ umbrella order. (e) Kagome $\sqrt{3}\times \sqrt{3}$ umbrella order. The convention is the same as Fig.~\ref{fig:RMO_hexagonal}.}
\label{fig:RMO_kagome}
\end{figure}

We use the following convention for the kagome lattice. The primitive lattice vectors: $a_1 = (1,0)$,$a_2 = (-\frac{1}{2},\frac{\sqrt{3}}{2})$. $a_3 = -a_1-a_2 = (-\frac{1}{2},-\frac{\sqrt{3}}{2})$. The origin is at the center of a hexagon. The vector $a_1/2$, $a_2/2$, and $a_3/2$ point to an A, B, and C site, respectively.

\begin{enumerate}
\item Kagome octahedral order (Fig.~\ref{fig:RMO_kagome}a). The spin configuration is given by:
\begin{align}
S(r) = \left\{ \begin{array}{cc}
N\hat{e}^{\snum1} e^{iQ_1\cdot r} & (r\in A)\\
N\hat{e}^{\snum2} e^{iQ_2\cdot r} & (r\in B)\\
N\hat{e}^{\snum3} e^{iQ_3\cdot r} & (r\in C)
\end{array}\right. .
\end{align}
$Q_{1,2,3}$ are the three wave vectors associated with the M points of the Brillouin zone. $N$ is a constant. The lattice translation $T_1$ maps $S(r)$ to:
\begin{align}
S'(r) = \left\{ \begin{array}{cc}
N\hat{e}^{\snum1} e^{iQ_1\cdot r} & (r\in A)\\
-N\hat{e}^{\snum2} e^{iQ_2\cdot r} & (r\in B)\\
-N\hat{e}^{\snum3} e^{iQ_3\cdot r} & (r\in C)
\end{array}\right. .
\end{align}
The spin $R(\hat{e}^{\snum1},\pi)$ rotation restores the magnetic order. We thus find a translation generator $(T_1,R(\hat{e}^{\snum1},\pi))$. Likewise, we find the other translation generator is $(T_2,R(\hat{e}^{\snum2},\pi))$. The lattice $C^z_2$ rotation about the origin maps a sublattice onto itself and $r\to -r$. It is easy to see that $S(r)$ is invariant under this operation. Therefore, $(C^z_2,1)$ is in the SSG.

\item Kagome cuboc1 order (Fig.~\ref{fig:RMO_kagome}b). The spin configuration is given by:
\begin{align}
S(r) = \left\{ \begin{array}{cc}
N (\hat{e}^{\snum2} e^{iQ_2\cdot (r-a_1/2)} + \hat{e}^{\snum3} e^{iQ_3 \cdot (r-a_1/2)}) & (r\in A)\\
N (\hat{e}^{\snum3} e^{iQ_3\cdot (r-a_2/2)} + \hat{e}^{\snum1} e^{iQ_1 \cdot (r-a_2/2)}) & (r\in B)\\
N (\hat{e}^{\snum1} e^{iQ_1\cdot (r-a_3/2)} + \hat{e}^{\snum2} e^{iQ_2 \cdot (r-a_3/2)}) & (r\in C)
\end{array}\right. .
\end{align}
The lattice translation $T_1$ maps $S(r)$ to:
\begin{align}
S'(r) = \left\{ \begin{array}{cc}
N (-\hat{e}^{\snum2} e^{iQ_2\cdot (r-a_1/2)} - \hat{e}^{\snum3} e^{iQ_3 \cdot (r-a_1/2)}) & (r\in A)\\
N (-\hat{e}^{\snum3} e^{iQ_3\cdot (r-a_2/2)} + \hat{e}^{\snum1} e^{iQ_1 \cdot (r-a_2/2)}) & (r\in B)\\
N (\hat{e}^{\snum1} e^{iQ_1\cdot (r-a_3/2)} - \hat{e}^{\snum2} e^{iQ_2 \cdot (r-a_3/2)}) & (r\in C)
\end{array}\right. .
\end{align}
$R(\hat{e}^{\snum1},\pi)$ restores the magnetic order. We thus obtain the translation generator $(T_1,R(\hat{e}^{\snum1},\pi))$. The other translation generator is $(T_2,R(\hat{e}^{\snum2},\pi))$. The lattice $C^z_2$ operation about the origin maps $S(r)$ to:
\begin{align}
S'(r) &=\left\{ \begin{array}{cc}
N (\hat{e}^{\snum2} e^{iQ_2\cdot (-r-a_1/2)} + \hat{e}^{\snum3} e^{iQ_3 \cdot (-r-a_1/2)}) & (r\in A)\\
N (\hat{e}^{\snum3} e^{iQ_3\cdot (-r-a_2/2)} + \hat{e}^{\snum1} e^{iQ_1 \cdot (-r-a_2/2)}) & (r\in B)\\
N (\hat{e}^{\snum1} e^{iQ_1\cdot (-r-a_3/2)} + \hat{e}^{\snum2} e^{iQ_2 \cdot (-r-a_3/2)}) & (r\in C)
\end{array}\right. 
\nonumber\\
& =  \left\{ \begin{array}{cc}
-N (\hat{e}^{\snum2} e^{iQ_2\cdot (r-a_1/2)} + \hat{e}^{\snum3} e^{iQ_3 \cdot (r-a_1/2)}) & (r\in A)\\
-N (\hat{e}^{\snum3} e^{iQ_3\cdot (r-a_2/2)} + \hat{e}^{\snum1} e^{iQ_1 \cdot (r-a_2/2)}) & (r\in B)\\
-N (\hat{e}^{\snum1} e^{iQ_1\cdot (r-a_3/2)} + \hat{e}^{\snum2} e^{iQ_2 \cdot (r-a_3/2)}) & (r\in C)
\end{array}\right. .
\end{align}
In the second line, we have used the fact that $\exp(iQ_{1}\cdot a_{2,3}) = -1$, and that $\exp(iQ_1\cdot (r-a_{2,3})) = \exp(-iQ_1\cdot (r-a_{2,3}))$, etc. The magnetic order is restored by $\Theta$, which yields the SSG element $(C^z_2,\Theta)$.

\item Kagome cuboc2 order (Fig.~\ref{fig:RMO_kagome}c). The spin configuration is given by:
\begin{align}
S(r) = \left\{ \begin{array}{cc}
N (\hat{e}^{\snum2} e^{iQ_2\cdot (r-a_1/2)} + \hat{e}^{\snum3} e^{iQ_3 \cdot (r-a_1/2)}) & (r\in A)\\
N (-\hat{e}^{\snum3} e^{iQ_3\cdot (r-a_2/2)} + \hat{e}^{\snum1} e^{iQ_1 \cdot (r-a_2/2)}) & (r\in B)\\
N (-\hat{e}^{\snum1} e^{iQ_1\cdot (r-a_3/2)} - \hat{e}^{\snum2} e^{iQ_2 \cdot (r-a_3/2)}) & (r\in C)
\end{array}\right. .
\end{align}
Note the magnetic structure within each sublattice is similar to that of the cuboc1. Therefore, the SSG translations and $C^z_2$ rotations are similar to that of cuboc1. 

\item Kagome $q=0$ order (Fig.~\ref{fig:RMO_kagome}d). The spin configuration is given by:
\begin{align}
S(r) = \left\{ \begin{array}{cc}
N\hat{e}^{\snum1} + M\hat{e}^{\snum3} & (r\in A)\\
N(-\frac{1}{2}\hat{e}^{\snum1}+\frac{\sqrt{3}}{2}\hat{e}^{\snum2}) + M\hat{e}^{\snum3} & (r\in B)\\
N(-\frac{1}{2}\hat{e}^{\snum1}-\frac{\sqrt{3}}{2}\hat{e}^{\snum2}) + M\hat{e}^{\snum3} & (r\in C)
\end{array}\right. .
\end{align}
As the magnetic order is uniform within each sublattice, the SSG translations and $C^z_2$ rotations do not require any accompanying spin operations.

\item Kagome $\sqrt{3}\times \sqrt{3}$ order (Fig.~\ref{fig:RMO_kagome}e). The spin configuration is given by:
\begin{align}
S(r) &= \left\{ \begin{array}{cc}
N(\hat{e}^{\snum1} \cos(Q \cdot (r - \frac{a_1}{2})) + \hat{e}^{\snum2} \sin(Q\cdot (r - \frac{a_1}{2}))) + M \hat{e}^{\snum3} & (r\in A)\\
N(\hat{e}^{\snum1} \cos(Q \cdot (r - \frac{a_2}{2})) + \hat{e}^{\snum2} \sin(Q\cdot (r - \frac{a_2}{2}))) + M \hat{e}^{\snum3} & (r\in B)\\
N(\hat{e}^{\snum1} \cos(Q \cdot (r - \frac{a_3}{2})) + \hat{e}^{\snum2} \sin(Q\cdot (r - \frac{a_3}{2}))) + M \hat{e}^{\snum3} & (r\in C)
\end{array}\right. 
\nonumber\\
& = \left\{ \begin{array}{cc}
N(\frac{\hat{e}^{\snum1}-i\hat{e}^{\snum2}}{2} e^{iQ \cdot (r - \frac{a_1}{2})} + c.c.) + M \hat{e}^{\snum3} & (r\in A)\\
N(\frac{\hat{e}^{\snum1}-i\hat{e}^{\snum2}}{2} e^{iQ \cdot (r - \frac{a_2}{2})} + c.c.) + M \hat{e}^{\snum3} & (r\in B)\\
N(\frac{\hat{e}^{\snum1}-i\hat{e}^{\snum2}}{2} e^{iQ \cdot (r - \frac{a_3}{2})} + c.c.) + M \hat{e}^{\snum3} & (r\in C)
\end{array}\right. .
\end{align}
$Q$ is at the K point of the Brillouin zone: $Q = \frac{4\pi}{3}(1,0)$. Each sublattice hosts a hexagonal F-umbrella state. Therefore, the SSG translations are similar to that of the hexagonal F-umbrella state. The lattice $C^z_2$ rotation about the origin maps $r$ to $-r$ within each sublattice:
\begin{align}
S'(r) &=\left\{ \begin{array}{cc}
N(\frac{\hat{e}^{\snum1}-i\hat{e}^{\snum2}}{2} e^{-iQ \cdot (r+\frac{a_1}{2})} + c.c.) + M \hat{e}^{\snum3} & (r\in A)\\
N(\frac{\hat{e}^{\snum1}-i\hat{e}^{\snum2}}{2} e^{-iQ \cdot (r+\frac{a_2}{2})} + c.c.) + M \hat{e}^{\snum3} & (r\in B)\\
N(\frac{\hat{e}^{\snum1}-i\hat{e}^{\snum2}}{2} e^{-iQ \cdot (r+ \frac{a_3}{2})} + c.c.) + M \hat{e}^{\snum3} & (r\in C)
\end{array}\right.  
\nonumber\\
& = \left\{ \begin{array}{cc}
N(\frac{\hat{e}^{\snum1}-i\hat{e}^{\snum2}}{2} e^{i\frac{2\pi}{3}} e^{-iQ \cdot (r-\frac{a_1}{2})} + c.c.) + M \hat{e}^{\snum3} & (r\in A)\\
N(\frac{\hat{e}^{\snum1}-i\hat{e}^{\snum2}}{2} e^{i\frac{2\pi}{3}} e^{-iQ \cdot (r-\frac{a_2}{2})} + c.c.) + M \hat{e}^{\snum3} & (r\in B)\\
N(\frac{\hat{e}^{\snum1}-i\hat{e}^{\snum2}}{2} e^{i\frac{2\pi}{3}} e^{-iQ \cdot (r- \frac{a_3}{2})} + c.c.) + M \hat{e}^{\snum3} & (r\in C)
\end{array}\right. .
\end{align}
In the second line, we have used the fact that $\exp(-iQ\cdot a_{1,2,3}) = e^{i\frac{2\pi}{3}}$. The spin space operation $\Theta R(\frac{\sqrt{3}}{2}\hat{e}^{\snum1}-\frac{1}{2}\hat{e}^{\snum2},\pi)$ restores the magnetic order.
 \end{enumerate}

\subsection{Square lattice} 
 
\begin{figure}
\centering
\includegraphics[width = \textwidth]{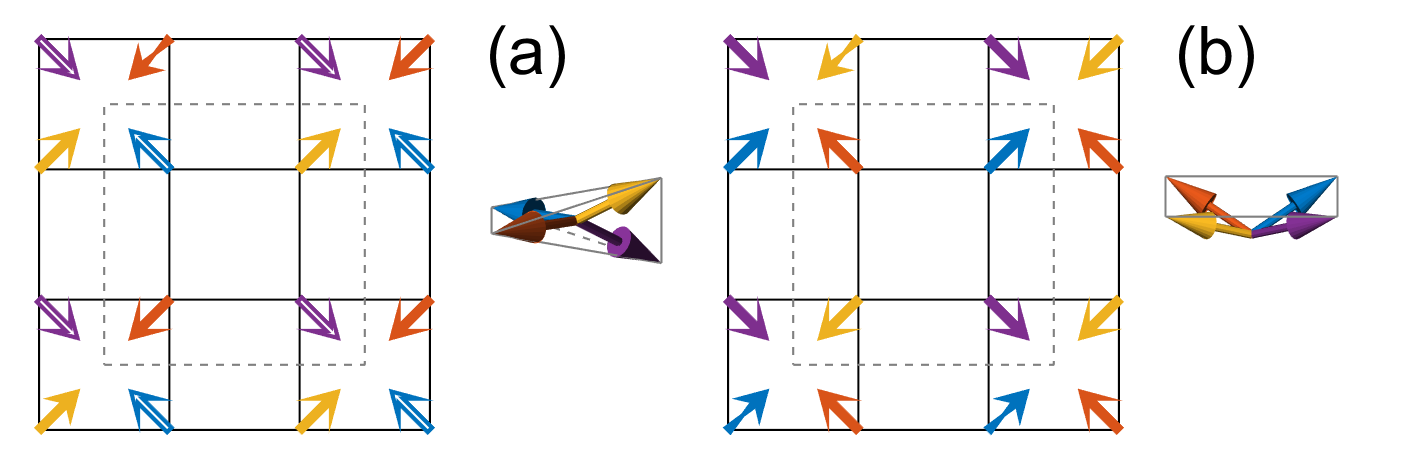}
\caption{(a) Square lattice tetrahedral umbrella order. (b) Square lattice umbrella order. The convention is the same as Fig.~\ref{fig:RMO_hexagonal}.}
\label{fig:RMO_square}
\end{figure}

We use the following convention for the square lattice. The primitive lattice vectors $a_1 = (1,0)$, and $a_2 = (0,1)$. $Q_{1,2}$ are at the X points of the first Brillouin zone: $Q_1 = (\pi,0)$, $Q_2 = (0,\pi)$. $Q_3 = Q_1+Q_2 = (\pi,\pi)$ is at the M point. The origin is at a site.

\begin{enumerate}

\item Square lattice tetrahedral umbrella order (Fig.~\ref{fig:RMO_square}a). The spin configuration is given by:
\begin{align}
S(r) = N(\hat{e}^{\snum1} e^{iQ_1\cdot r} + \hat{e}^{\snum2} e^{iQ_2 \cdot r})+ N'\hat{e}^{\snum3} e^{iQ_3\cdot r}.
\end{align}
$N$ and $N'$ are constants. The lattice translation $T_1$ maps the spin configuration to:
\begin{align}
S'(r) = N( -\hat{e}^{\snum1} e^{iQ_1\cdot r} + \hat{e}^{\snum2} e^{iQ_2 \cdot r}) - N' \hat{e}^{\snum3} e^{iQ_3\cdot r}.
\end{align}
The spin $R(\hat{e}^{\snum2},\pi)$ rotation restores the magnetic order, yielding the SSG translation generator $(T_1,R(\hat{e}^{\snum2},\pi))$. The other SSG translation generator $(T_2,R(\hat{e}^{\snum1},\pi))$ is obtained in the same vein. The lattice $C^z_2$ rotation about the origin leaves $S(r)$ invariant:
\begin{align}
S'(r) &=  N(\hat{e}^{\snum1} e^{-iQ_1\cdot r} + \hat{e}^{\snum2} e^{-iQ_2 \cdot r})+ N'\hat{e}^{\snum3} e^{-iQ_3\cdot r}
\nonumber\\
&= N(\hat{e}^{\snum1} e^{iQ_1\cdot r} + \hat{e}^{\snum2} e^{iQ_2 \cdot r})+ N'\hat{e}^{\snum3} e^{iQ_3\cdot r} = S(r).
\end{align}
In the second line, we have used $\exp(iQ_{1,2,3}\cdot r) = \exp(-iQ_{1,2,3}\cdot r)$. This operation corresponding to the SSG element $(C^z_2,1)$.

\item Square lattice umbrella order (Fig.~\ref{fig:RMO_square}b). The spin configuration is given by:
\begin{align}
S(r) = N( \hat{e}^{\snum1} e^{iQ_1\cdot r} + \hat{e}^{\snum2} e^{iQ_2 \cdot r} )+ M\hat{e}^{\snum3} .
\end{align}
The lattice translation $T_1$ maps $S(r)$ to:
\begin{align}
S'(r) = N(-\hat{e}^{\snum1} e^{iQ_1\cdot r} + \hat{e}^{\snum2} e^{iQ_2 \cdot r} )+ M\hat{e}^{\snum3},
\end{align}
which is restored by $\Theta R(\hat{e}^{\snum1},\pi)$. This symmetry operation corresponds to $(T_1,\Theta R(\hat{e}^{\snum1},\pi))$. The other SSG translation is $(T_2,\Theta R(\hat{e}^{\snum2}))$. As the lattice $C^z_2$ leaves $S(r)$ invariant, the SSG $C^z_2$ rotation is simply $(C^z_2,1)$.
\end{enumerate}

\subsection{Nonreciprocity and stability}

We apply the reciprocity and stability criteria to these magnetic orders. We find that:
\begin{itemize}
\item The spin waves in hexagonal lattice F-umbrella,  honeycomb cubic, kagome cuboc 1\&2, kagome $\sqrt{3}\times \sqrt{3}$ umbrella orders are non-reciprocal in zero electric field because they lack symmetries protect the spectral reciprocity in zero electric field.
\item The spin waves in square lattice umbrella order is reciprocal in electric field because the anti-unitary translation symmetry protect the reciprocity.
\item The hexagonal F-umbrella, kagome $q=0$ umbrella, kagome $\sqrt{3}\times\sqrt{3}$ umbrella states are unstable in electric field because the translation symmetry permits a linear gradient term in potential energy, leading to the magnetic order's instability.
\item The spin waves in hexagonal lattice tetrahedral, honeycomb tetrahedral, kagome otcahedral, and square lattice tetrahedral order are reciprocal in zero electric field, and should exhibit non-reciprocity in electric field.
\end{itemize}

\subsection{Instability of kagome $q=0$ umbrella order in electric field}

\begin{figure}
\centering
\includegraphics[width = 0.8\textwidth]{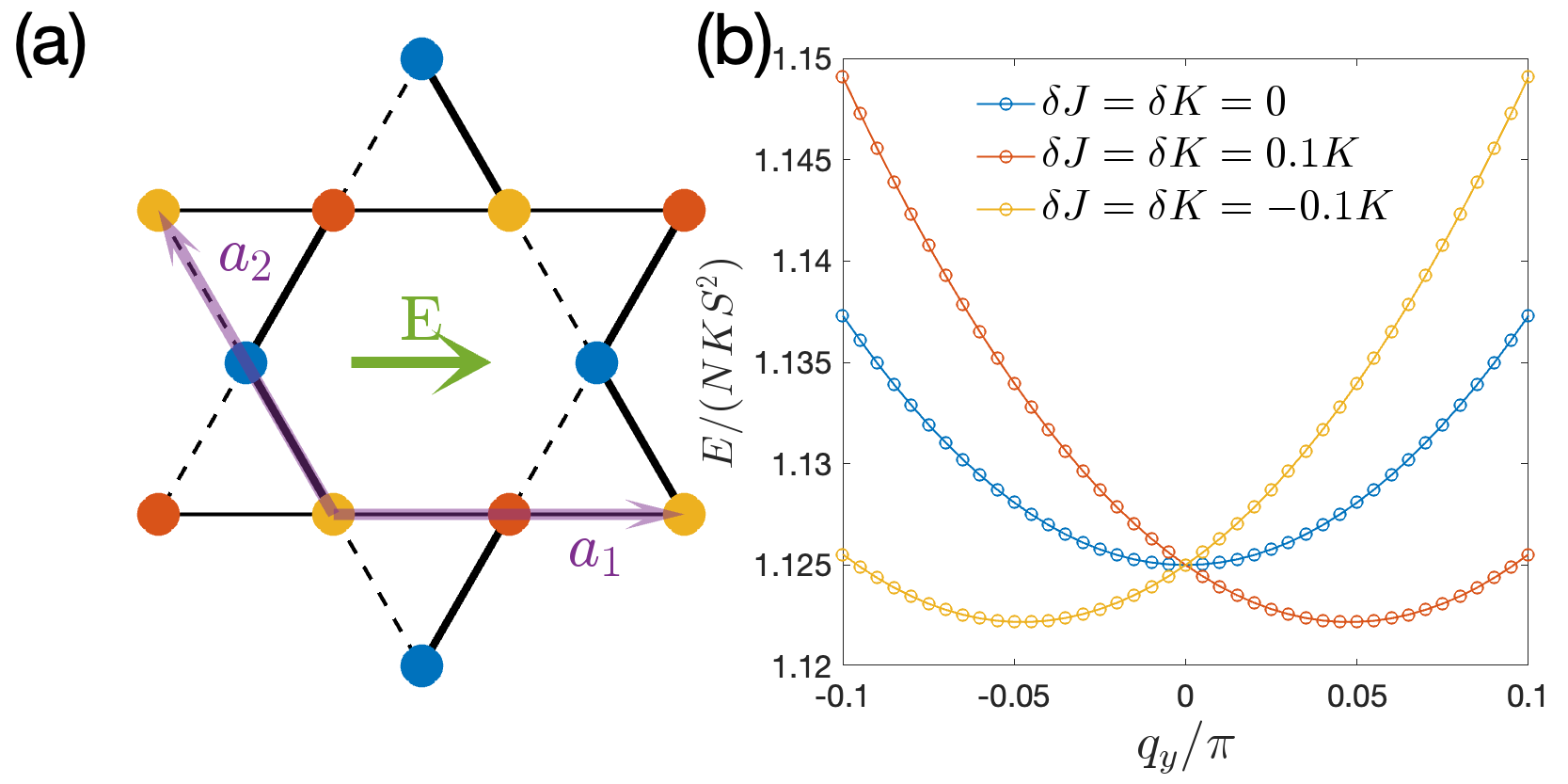}
\caption{(a) Kagome lattice. A, B, C sites are colored in red, blue, and yellow. Thick solid lines (thin dashed lines) denote bonds where the exchange interactions are enhanced (weakened) in the presence of an electric field (green arrow) along the lattice horizontal direction. $a_{1,2}$ are the primitive lattice vectors. (b) Energy density of a spiral state as a function of wave vector $q$ for a bilinear-biquadratic model. $q$ is in the lattice vertical direction. $\theta = \pi/4$. We set $K>0$ and $J$ according to Eq.~\eqref{eq:theta_condition}.}
\label{fig:kagome_umbrella}
\end{figure}

We have argued by symmetry that, if the translation symmetry permits the linear gradient term in the potential energy, the magnetic order is generically unstable toward spiral formation in electric field. $q=0$ umbrella order falls into this category. In this section, we show that, by a variational calculation, such instability indeed occurs.

To this end, we construct a Hamiltonian whose ground state is the $q=0$ umbrella order (Fig.~\ref{fig:kagome_umbrella}a):
\begin{align}
\hat{n}_r = \left\{ \begin{array}{cc}
(\sin\theta,0,\cos\theta) & (r\in A) \\
(-\frac{1}{2}\sin\theta,\frac{\sqrt{3}}{2}\sin\theta,\cos\theta) & (r\in B) \\
(-\frac{1}{2}\sin\theta,-\frac{\sqrt{3}}{2}\sin\theta,\cos\theta) & (r\in C)
\end{array}\right. .
\end{align}
$\theta \in (0,\pi/2)$ is an angle parameter. $\hat{n}_r$ is the unit vector parametrizing the orientation of the spin. In particular, the dot product of the unit vectors from two neighboring sites:
\begin{align}
\hat{n}(r) \cdot \hat{n}(r') = 1-\frac{3}{2}\sin^2\theta. 
\end{align}
We consider a classical Heisenberg bilinear-biquadratic model on kagome:
\begin{align}
H = \sum_{\langle rr'\rangle} J S(r)\cdot S(r') + \frac{K}{S^2} [S(r)\cdot S(r')]^2 = \sum_{\langle rr'\rangle} JS^2 \hat{n}(r)\cdot \hat{n}(r') + KS^2 [\hat{n}(r)\cdot \hat{n}(r')]^2 = \sum_{\langle rr'\rangle} H_{rr'}.
\end{align}
$H_{rr'}$ is the local Hamiltonian on a nearest neighbor bond $rr'$. Minimizing $H_{rr'}$ with respect to $\hat{n}(r) \cdot \hat{n}(r')$ yields:
\begin{align}
\hat{n}(r) \cdot \hat{n}(r') = -\frac{J}{2K} = 1- \frac{3}{2}\sin^2\theta.
\label{eq:theta_condition}
\end{align}
We demand that $\hat{n}(r)\cdot \hat{n}(r')$ is the prescribed value. With this condition, the $q=0$ umbrella order is a ground state of the Hamiltonian $H$ in that all $H_{rr'}$ attain their minimum. It is quite clear that the ground state is not \emph{unique}: there are infinitely many discrete families of ground states. The $q=0$ umbrella order is just one of them. However, it is not an issue for our purpose in that our interest lies in the \emph{local instability} of the $q=0$ umbrella order.

As the electric field breaks the lattice rotational symmetry and inversion symmetry, the exchange interactions on different bonds are no longer equal. For concreteness, we set the electric field $\parallel x$ axis (Fig.~\ref{fig:kagome_umbrella}a). We assume the exchange parameters on the thick solid bonds are enhanced, i.e. $J+\delta J,K+\delta K$, whereas the exchange parameters on the thin dashed bonds are reduced, i.e. $J-\delta J,K-\delta K$. The rest remain the same.

We use the following trial ground state: 
\begin{align}
\hat{n}_r = \left\{ \begin{array}{cc}
(\sin\theta \cos(q\cdot r),\sin\theta \sin(q\cdot r),\cos\theta) & (r\in A) \\
(\sin\theta \cos(q\cdot r + \frac{2\pi}{3}),\sin\theta \sin(q\cdot r+\frac{2\pi}{3}),\cos\theta) & (r\in B) \\
(\sin\theta \cos(q\cdot r - \frac{2\pi}{3}),\sin\theta \sin(q\cdot r-\frac{2\pi}{3}),\cos\theta) & (r\in C)
\end{array}\right. .
\end{align}
The above describes a single spiral state with wave vector $q$. In particular, the above state at $q=0$ reduces to the $q=0$ umbrella order. The energy density of this state is given by:
\begin{subequations}
\begin{align}
\frac{E(q)}{N} = \frac{\epsilon_{AB+} + \epsilon_{AB-} + \epsilon_{CA+} + \epsilon_{CA-} + \epsilon_{BC+} + \epsilon_{BC-}}{3},
\end{align}
where the bond energy:
\begin{align}
\epsilon_{AB+} &= (J\mp\delta J)S^2 [\cos^2\theta+\sin^2\theta \cos(\frac{2\pi}{3}\pm\frac{q\cdot a_3}{2})]+(K\mp\delta K)S^4[\cos^2\theta+\sin^2\theta \cos(\frac{2\pi}{3}\pm\frac{q\cdot a_3}{2})]^2
\\
\epsilon_{CA\pm} &= (J\pm\delta J)S^2 [\cos^2\theta+\sin^2\theta \cos(\frac{2\pi}{3} \pm \frac{q\cdot a_2}{2})]+ (K\pm\delta K)S^4 [\cos^2\theta+\sin^2\theta \cos(\frac{2\pi}{3} \pm \frac{q\cdot a_2}{2})]^2;
\\
\epsilon_{BC\pm} &= JS^2[\cos^2\theta+\sin^2\theta \cos(\frac{2\pi}{3}\pm \frac{q\cdot a_1}{2})]+KS^4[\cos^2\theta+\sin^2\theta \cos(\frac{2\pi}{3}\pm\frac{q\cdot a_1}{2})]^2.
\end{align}
\end{subequations}
$a_{1,2}$ are the two primitive lattice vectors; $a_3 = -a_1-a_2$ (Fig.~\ref{fig:kagome_umbrella}a).

Fig.~\ref{fig:kagome_umbrella}b shows the energy density of the spiral state for $q$ along the lattice $y$ direction for $\theta = \pi/4$. We find that, in the absence of any modulation ($\delta J = \delta K = 0$), the energy minimum is at $q=0$, i.e. the $q=0$ umbrella order. However, when the electric-field induced modulation is present, the minimum shifts to finite $q$. In particular, the minimum is at $q>0$ ($q<0$) for $\delta J=\delta K>0$ ($<0$). This simple model illustrates that the kagome $q=0$ umbrella order is indeed unstable in the presence of electric field. Finally, we note that, for special parameters $\delta J/J = \delta K/K$, the $q=0$ state is stable since the condition Eq.~\eqref{eq:theta_condition} is still met. However, this stability is accidental in that it is not protected by symmetry. 

\section{Goldstone modes in hexagonal tetrahedral order}

In the main text, we have constructed the following field theory Lagrangian for the Goldstone modes in the hexagonal tetrahedral order:
\begin{subequations}
\label{eq:goldstone_lagrangian}
\begin{align}
\mathcal{L} = m^\alpha \partial_t \theta^\alpha  - \mathcal{U}_0 - \mathcal{U}_1;
\\
\mathcal{U}_0 = \frac{\rho_\parallel}{2} \partial_\parallel \theta^\alpha \partial_\parallel \theta^\alpha + \frac{\rho_\perp}{2} \partial_\perp \theta^\alpha \partial_\perp \theta^\alpha + \frac{m^\alpha m^\alpha}{2\chi};
\\
\mathcal{U}_1 = -\zeta_1 E_\perp m^\alpha \partial_\parallel \theta^\alpha -\zeta_2 E_\parallel m^\alpha \partial_\perp \theta^\alpha .
\end{align}
\end{subequations}
In this section, we detail the analysis of Eq.~\eqref{eq:goldstone_lagrangian}.

Applying the Euler-Lagrangian equation to Eq.~\eqref{eq:goldstone_lagrangian}, we obtain the spin wave equation:
\begin{subequations}
\begin{align}
(\partial_t + \zeta_1 E_\perp \partial_\parallel  + \zeta_2 E_\parallel \partial_\perp) \theta^\alpha = \frac{m^\alpha}{\chi};
\\
(\partial_t + \zeta_1 E_\perp \partial_\parallel + \zeta_2 E_\parallel \partial_\perp) m^\alpha = (\rho_\parallel \partial^2_\parallel  + \rho_\perp \partial^2_\perp) \theta^\alpha.
\end{align}
\end{subequations}
Note fields with different label $\alpha$ are decoupled. Inserting the plane wave solution: $\theta^\alpha (x,t) = \theta^\alpha e^{i(q\cdot x-\omega t)}$, $m^\alpha (x,t) = m^\alpha e^{i(q\cdot x-\omega t)}$ into the equation, we find:
\begin{align}
i(-\omega + \zeta_1 E_\perp q_\parallel + \zeta_2 E_\parallel q_\perp )\theta^\alpha = \frac{m^\alpha}{\chi};
\quad
i(-\omega + \zeta_1 E_\perp q_\parallel + \zeta_2 E_\parallel q_\perp )m^\alpha = -(\rho_\parallel q^2_\parallel + \rho_\perp q^2_\perp) \theta^\alpha;
\end{align}
Solving the above for $\omega$, we obtain the dispersion relation:
\begin{align}
\omega (q) = \zeta_1 E_\perp q_\parallel + \zeta_2 E_\parallel q_\perp + \sqrt{\frac{\rho_\parallel q^2_\parallel + \rho_\perp q^2_\perp}{\chi}}.
\end{align}

The positive semidefinite property of the potential energy $\mathcal{U}_0+\mathcal{U}_1\ge 0$ requires:
\begin{align}
\begin{pmatrix}
\rho_\parallel q^2_\parallel + \rho_\perp q^2_\perp & -i(\zeta_1 E_\perp q_\parallel + \zeta_2 E_\parallel q_\perp)  \\
-i(\zeta_1 E_\perp q_\parallel + \zeta_2 E_\parallel q_\perp)  & \frac{1}{\chi} 
\end{pmatrix} \succeq 0,
\end{align} 
for any $q$. Since the diagonal entires are both positive semidefinite, the relevant constraint is:
\begin{align}
\frac{\rho_\parallel q^2_\parallel + \rho_\perp q^2_\perp}{\chi} \ge (\zeta_1 E_\perp q_\parallel + \zeta_2 E_\parallel q_\perp)^2.
\end{align}
We may view the above as an quadratic form in $q_{\parallel,\perp}$:
\begin{align}
f(q_\parallel,q_\perp) = [\frac{\rho_\parallel}{\chi}-(\zeta_1 E_\perp)^2]q^2_\parallel + [\frac{\rho_\perp}{\chi}-(\zeta_2 E_\parallel)^2]q^2_\perp - 2\zeta_1\zeta_2 E_\perp E_\parallel q_\parallel q_\perp \ge 0.
\end{align}
The above condition holds for any $q_\parallel,q_\perp$ if and only if:
\begin{align}
[\frac{\rho_\parallel}{\chi}-(\zeta_1 E_\perp)^2][\frac{\rho_\perp}{\chi}-(\zeta_2 E_\parallel)^2]\ge (\zeta_1 E_\perp)^2 (\zeta_2 E_\parallel)^2.
\end{align}
Therefore, the positive semidefiniteness of the potential energy sets an upper bound on the electric field; beyond this bound, the tetrahedral order loses stability if the dielectric breakdown has not occurred already. We note this instability is different from those discussed in the main text, which is about the instability at infinitesimal electric field.

\section{Adiabatic spin wave theory for spin waves in electric field}

We apply the adiabatic spin wave theory~\cite{Niu1998,Halilov1998,Zhou2025} to magnetic insulators in the electric field. While we consider the classical Kondo model~\cite{Martin2008,Akagi2010,Kato2010} for concreteness, the formalism can be extended to other models for magnetic insulators.

\subsection{Lagrangian}

We consider the classical Kondo model subject to a uniform electric field:
\begin{align}
H = \sum_{\mathbf{k}s} \epsilon_{\mathbf{k}-e\mathbf{A}} c^\dagger_{\bm{k}s} c^{\phantom\dagger}_{\bm{k}s} - JS \sum_{r} \hat{n}_r\cdot \bm{m}_r.
\end{align}
Here, we employ the velocity gauge. $\bm{A}$ is a uniform vector potential; $\bm{E} = -\dot{\bm{A}}$ is the electric field. $e = -|e|$ is the electric charge. $\epsilon_{\bm{k}}$ is the dispersion relation of the electron.  $J>0$ is the ferromagnetic exchange constant. $S$ is the spin length. $\hat{n}_r$ is a unit vector in the spin space that describes the direction of the classical spin at site $r$. $s = \uparrow,\downarrow$ labels the electron spin.  $\bm{m}_{r} =  \sum_{ss'} \bm{\sigma}_{ss'} c^\dagger_{rs} c_{rs'}$ is the electron magnetization at $r$. 

The dynamics of this hybrid classical-quantum system is captured by the following Lagrangian:
\begin{align}
L = S\sum_r \mathcal{A}_r \cdot \dot{\hat{n}}_r + \langle \Psi|\dot{\Psi}\rangle - \langle \Psi|H|\Psi\rangle.
\label{eq:lagrangian}
\end{align}
$\mathcal{A}_r = \mathcal{A}(\hat{n}_r)$ is a Berry connection in the spin space; $\nabla_n\times \mathcal{A} = -S\hat{n}$. The first term may be viewed as the kinetic term for the spin. $|\Psi\rangle$ is the many-body state of the electrons. 

We derive an effective Lagrangian for the spins. We assume that the bandwidth of the spin waves and $eEa$, where $a$ is the lattice constant, are both much smaller than the electron particle-hole excitation gap. In this case, we may invoke the adiabatic approximation:
\begin{align}
|\Psi\rangle = |\Psi_G(\{\hat{n}\},\bm{A})\rangle,
\end{align}
where the right hand side is the instantaneous ground state of the Hamiltonian $H$ for given spin configuration $\{\hat{n}\}$ and the vector potential $\bm{A}$. Substituting the above ans\"{a}tz into Eq.~\eqref{eq:lagrangian}, we obtain the following effective Lagrangian for the spins:
\begin{align}
L = \sum_r \mathcal{A}'_r \cdot \dot{\hat{n}}_r - \mathcal{E}_G  + \bm{P} \cdot \bm{E}. 
\label{eq:L_spin}
\end{align}
Here, $\mathcal{A}'_r$ accounts for the electron Berry connection contribution:
\begin{align}
\mathcal{A}'_r = \mathcal{A}_r + i\langle\Psi_G | \nabla_{\hat{n}_r}\Psi_G \rangle.
\end{align}
$\mathcal{E}_G$ is the instantaneous ground state energy of the electrons. The equilibrium spin configurations correspond to the minima of $\mathcal{E}_G$. The polarization $\bm{P}$ is given by~\cite{Resta2007}:
\begin{align}
\bm{P}=  -i\langle\Psi_G | \nabla_{\bm{A}}\Psi_G \rangle.
\end{align}

We are interested in the linear spin waves about the equilibrium spin configuration. It is convenient to rotate the local spin frames such that the local $\hat{e}^{\snum3}_{r}$ axis coincides with $\hat{n}_{r}$ at equilibrium. We write:
\begin{align}
\hat{n}_r \approx u^{\snum1}_r  \hat{e}^{\snum1}_r + u^{\snum2}_r \hat{e}^{\snum2}_r + (1-\frac{u^2_r}{2})\hat{e}^{\snum3}_r.
\end{align}
$\hat{e}^{\snum{1},\snum{2},\snum{3}}$ form a right-hand spin frame at site $r$. $u^{\snum{1},\snum{2}}_r$ parametrizes the deviations from ground state. Expanding Eq.~\eqref{eq:L_spin} to quadratic order in $u^{\snum{1},\snum{2}}_r$ and first order in $\bm{A}$, we obtain:
\begin{align}
L = \frac{1}{2}\sum_{\alpha r,\beta r'}(\Omega^{\alpha\beta}_{rr'} u^\alpha_{r} \dot{u}^\beta_{r'} - \Phi^{\alpha\beta}_{rr'}u^\alpha_r u^\beta_{r'})  + \bm{E} \cdot \bm{P}.
\label{eq:L_spin_linear}
\end{align}

The first term in Eq.~\eqref{eq:L_spin_linear} is the kinetic term. $\Omega$ is a real skew-symmetric matrix. It is given by:
\begin{align}
\Omega^{\alpha\beta}_{rr'} =  -S\epsilon^{\alpha\beta}\delta_{rr'} + i (\langle \frac{\partial\Psi_G}{\partial u^\alpha_r}| \frac{\partial \Psi_G}{\partial u^\beta_{r'}}\rangle - \langle \frac{\partial\Psi_G}{\partial u^\beta_{r'}}| \frac{\partial \Psi_G}{\partial u^\alpha_{r}}\rangle).
\end{align}
$\epsilon^{\alpha\beta}$ is the Levi-Civita tensor: $\epsilon^{12} = -\epsilon^{21} = 1$, $\epsilon^{11} = \epsilon^{22} = 0$. It encodes the Poisson bracket:
\begin{align}
\{u^\alpha_r, u^\beta_{r'}\} = (\Omega^{-1})^{\alpha\beta}_{rr'},
\end{align}
in the Hamiltonian formulation.

The second term is the energy term. There is no term linear in $u$ owing to the stationary condition at $u=0$. $\Phi^{ab}_{rr'}$ is the Hessian:
\begin{align}
\Phi^{ab}_{rr'} = \frac{\partial^2 \mathcal{E}_G}{\partial u^a_r \partial u^b_{r'}},
\end{align}
which is a real symmetric matrix. 

The last term couples the spin waves to the electric field. $\bm{P}$ is the electron polarization:
\begin{subequations}
\begin{align}
\bm{P} = \sum_{\alpha r} \bm{\Upsilon}^{\alpha}_{r} u^\alpha_r + \frac{1}{2}\sum_{\alpha r,\beta r'} \bm{Z}^{\alpha\beta}_{rr'} u^\alpha_r u^\beta_{r'}.
\end{align}
The coefficients can be viewed as analogues of the magnetoelectric coefficients:
\begin{align}
\bm{\Upsilon}^{\alpha}_{r} = i (\langle \frac{\partial\Psi_G}{\partial \bm{A} }| \frac{\Psi_G}{\partial u^\alpha_r}\rangle - \langle \frac{\partial\Psi_G}{\partial u^\alpha_r }| \frac{\Psi_G}{\partial \bm{A}}\rangle).
\\
\bm{Z}^{\alpha\beta}_{rr'}  = \frac{i}{2}\frac{\partial}{\partial u^\beta_{r'}}(\langle \frac{\partial\Psi_G}{\partial \bm{A} }| \frac{\Psi_G}{\partial u^\alpha_r}\rangle - \langle \frac{\partial\Psi_G}{\partial u^\alpha_r }| \frac{\Psi_G}{\partial \bm{A}}\rangle) + (u^\alpha_r \leftrightarrow u^\beta_{r'}).
\end{align}
\end{subequations}
In the case we are interested in, $\bm{\Upsilon} = 0$ by symmetry.

We note that the Lagrangian formally admits another term of the form $\bm{A}\cdot \bm{J}$, where
$\bm{J}$ may be interpreted as a current:
\begin{align}
\bm{J} = \frac{i}{2}\sum_{\alpha r,\beta r'} \frac{\partial}{\partial \bm{A}} (\langle \frac{\partial\Psi_G}{\partial u^\alpha_r}| \frac{\partial \Psi_G}{\partial u^\beta_{r'}}\rangle - \langle \frac{\partial\Psi_G}{\partial u^\beta_{r'}}| \frac{\partial \Psi_G}{\partial u^\alpha_{r}}\rangle) u^\alpha_r \dot{u}^\beta_{r'}.
\end{align}
However, as we shall see in Sec.~\ref{sec:no_current}, the derivative vanishes in the thermodynamic limit due to gauge invariance. We therefore drop this term in the Lagrangian at the outset.

\subsection{Equation of motion}

Returning to the linearized Lagrangian Eq.~\eqref{eq:L_spin_linear}. We first consider if the ground state is stable in the presence of $\bm{E}$. The total energy is given by:
\begin{align}
E_\mathrm{tot} = -\sum_{\alpha r} (\bm{E} \cdot \bm{\Upsilon}^{\alpha}_r) u^\alpha_r + \frac{1}{2}\sum_{\alpha r,\beta r'}(\Phi^{\alpha\beta}_{rr'} - \bm{E} \cdot \bm{Z}^{\alpha\beta}_{rr'}) u^\alpha_r u^\beta_{r'}.
\end{align}
If the linear coefficient $\bm{\Upsilon}$ is non-zero, then the electric field will induces a displacement in the equilibrium spin configuration, indicating that the ground state is unstable. Furthermore, the matrix:
\begin{align}
\Phi - \bm{E}\cdot \bm{Z} \succeq 0.
\end{align}
viz. positive semi-definite. This condition imposes a threshold for the electric field, beyond which the spin configuration loses stability. 

Provided that $\bm{\Upsilon} = 0$ and the stability condition holds, the equation of motion for $u$ reads:
\begin{align}
\sum_{\beta r'}\Omega^{\alpha \beta}_{rr'} \dot{u}^\beta_{r'} = \sum_{\beta r'} (\Phi^{\alpha\beta}_{rr'} - \bm{E}\cdot \bm{Z}^{\alpha\beta}_{rr'}) u^\beta_{r'}.
\end{align}
In the momentum space,
\begin{align}
\sum_{\beta l'} \Omega^{\alpha\beta}_{ll'}(q) \dot{u}^b_{l'}(q) = \sum_{\beta l'} (\Phi^{\alpha\beta}_{ll'}(q) - \bm{E}\cdot \bm{Z}^{\alpha\beta}_{ll'}(q)) u^\beta_{l'}(q).
\end{align}
Making the ans\"{a}tz $u^\beta_{l'}(q,t) = u^\beta_{l'}(q)e^{-i\omega(q) t}$,  the above equation reduces to the eigenvalue problem:
\begin{align}
i\Omega^{-1}(q)(\Phi(q)- \bm{E} \cdot \bm{Z}(q)) u(q) = \omega(q) u(q).
\end{align}
Provided that the matrix $K(q)- \bm{E}\cdot \bm{Z}(q) \succ 0$, the matrix on the left features pairs of positive and negative eigenvalues. The positive eigenvalues correspond to the frequency of the spin wave modes at wave vector $q$.

\subsection{Calculating Lagrangian parameters \label{sec:parameters}}

In this section, we provide a recipe for calculating the various parameters that appear in the effective Lagrangian assuming that the magnetic order at equilibrium features a translation symmetry.

\subsubsection{Hessian}

The Hessian is the second order derivative of the ground state energy with respect to $u$. The second order Rayleigh-Schr\"{o}dinger perturbation theory yields:
\begin{align}
\Phi^{\alpha\beta}_{rr'} = \langle \Psi_0|\frac{\partial^2 H}{\partial u^\alpha_r \partial u^\beta_{r'}}|\Psi_0\rangle - \sum_{n>0}\frac{\langle \Psi_0 |\frac{\partial H}{\partial u^\alpha_r}\Psi_n\rangle \langle \Psi_n| \frac{\partial H}{\partial u^\beta_{r'}} |\Psi_0\rangle+\langle \Psi_0 |\frac{\partial H}{\partial u^\beta_{r'}}|\Psi_n\rangle \langle \Psi_n| \frac{\partial H}{\partial u^\alpha_r} |\Psi_0\rangle}{E_n - E_0}.
\end{align}
Here, all Hamiltonian derivatives are carried out at $u=0$ and $\bm{A}=0$ unless stated otherwise. $E_n$ and $|\Psi_n\rangle$ are respectively the \emph{many-body} eigenvalue and eigenstate of the \emph{equilibrium} Hamiltonian $H_0$, i.e. the Hamiltonian with $u=0$ and $\bm{A} = 0$. $n=0$ corresponds to the ground state. The various derivatives of the Hamiltonian are given by:
\begin{align}
\frac{\partial H}{\partial u^\alpha_r} = -JS m^\alpha_r; \quad
\frac{\partial^2 H}{\partial u^\alpha_r \partial u^\beta_{r'}} = JS m^{\snum3}_r\delta_{\alpha\beta}\delta_{rr'}.
\quad (\alpha,\beta=1,2).
\end{align}
Here, $m^\alpha_i$ is the electron magnetization density operator on site $r$ in the local spin frame:
\begin{align}
m^\alpha_r = \sum_{ss'} (\hat{e}^{\alpha}_r \cdot \sigma)_{ss'} c^\dagger_{rs} c^{\phantom\dagger}_{rs'}.
\end{align}
Inserting the derivatives in, we obtain:
\begin{align}
\Phi^{\alpha\beta}_{rr'} = JS\langle \Psi_0|m^z_{r}|\Psi_0\rangle \delta_{\alpha\beta}\delta_{rr'} - (JS)^2\sum_{n>0}\frac{\langle \Psi_0 |m^\alpha_r|\Psi_n\rangle \langle \Psi_n|m^\beta_{r'}|\Psi_0\rangle+\langle \Psi_0 |m^\beta_{r'}|\Psi_n\rangle \langle \Psi_n|m^\alpha_{r}|\Psi_0\rangle}{E_n - E_0}.
\end{align}

The equilibrium spin configuration features a translation symmetry. We switch to the momentum space:
\begin{align}
\Phi^{\alpha\beta}_{ll'}(\bm{q}) = \frac{1}{N_c}\sum_{i\in l} \sum_{j \in l'} \Phi^{\alpha\beta}_{rr'} e^{-i \bm{q}\cdot (\bm{r}-\bm{r}')} =
JS m^{\snum3}_l\delta_{ll'}\delta_{\alpha\beta} - (JS)^2 \chi^{\alpha\beta}_{ll'}(\bm{q}),
\end{align} 
where $l,l'$ are the magnetic sublattice labels. $q$ is the wave vector. $N_c$ is the number of magnetic unit cells.  $m^{\snum3}_l$ is the magnetic moment in the local $e^{\snum3}$ direction on sublattice $l$:
\begin{align}
m^{\snum3}_l = \frac{1}{N_c} \langle \Psi_0|m^{\snum3}_l(\bm{q}=0)|\Psi_0\rangle.
\end{align}
Here, $m^\alpha_l(\bm{q})$ is the magnetization density in momentum space:
\begin{align}
m^\alpha_l (\bm{q}) = \sum_{r \in l} m^\alpha_r e^{-i\bm{q} \cdot \bm{r}}.
\end{align}
$\chi^{\alpha\beta}_{ll'}(\bm{q})$ is the susceptibility:
\begin{align}
\chi^{\alpha\beta}_{ll'}(\bm{q}) = \frac{1}{N_c} \sum_{n\neq 0}\frac{\langle \Psi_0 |m^\alpha_l (\bm{q})|\Psi_n\rangle \langle \Psi_n|m^\beta_{l'}(-\bm{q})|\Psi_0\rangle + \langle \Psi_0 |m^\beta_{l'} (-\bm{q})|\Psi_n\rangle \langle \Psi_n|m^\alpha_l (\bm{q})|\Psi_0\rangle}{E_n - E_0}.
\end{align}

We switch to the diagonal basis of the equilibrium Hamiltonian:
\begin{align}
H_0 = \sum_{\bm{k} \lambda} \epsilon_\lambda(\bm{k}) c^\dagger_\lambda(\bm{k}) c^{\phantom\dagger}_{\lambda}(\bm{k}),
\end{align}
where $\lambda$ is the band index. In this basis, the magnetization density operator reads:
\begin{align}
m^\alpha_l (\bm{q}) = \sum_{\bm{k}} \langle \bm{k}\lambda|m^\alpha_l(\bm{q})|\bm{k}+\bm{q}\mu\rangle c^\dagger_\lambda(\bm{k}) c^{\phantom\dagger}_\mu(\bm{k}+\bm{q}).
\end{align}
Here, $|\bm{k}\lambda\rangle$ is the \emph{one-body} state with momentum $\bm{k}$ and band index $\lambda$. The matrix element is given by:
\begin{align}
\langle \bm{k}\lambda|m^\alpha_l(q)|\bm{k}+\bm{q}\mu\rangle = \sum_{ls} (\hat{e}^{\alpha}_l\cdot \bm{\sigma})_{ss'} \langle \bm{k}\lambda|\bm{k},ls\rangle  \langle \bm{k}+\bm{q},ls'|\bm{k}+\bm{q}\mu \rangle.
\end{align}
Using these, we obtain the following expression for $m^{\snum3}_l$:
\begin{align}
m^{\snum3}_l = \frac{1}{N_c}\sum_{\bm{k}} \sum_{\lambda\in v}\langle k\lambda| m^{\snum3}_l(\bm{q}=0)|k\lambda\rangle.
\end{align}
Here, the summation of $\lambda$ is over the valence (``$v$") bands. The susceptibility $\chi^{\alpha\beta}_{ll'}(\bm{q})$ reads:
\begin{align}
\chi^{\alpha\beta}_{ll'}(\bm{q}) = \frac{1}{N_c}\sum_{\bm{k}} (\sum_{\lambda \in c} \sum_{\mu \in v} - \sum_{\lambda \in v} \sum_{\mu\in c}) \frac{\langle \bm{k}\mu|m^\alpha_l(\bm{q})|\bm{k}+\bm{q}\lambda\rangle \langle \bm{k}+\bm{q}\lambda |m^\beta_{l'}(-\bm{q})|\bm{k}\mu\rangle}{\epsilon_\lambda(\bm{k}+\bm{q})-\epsilon_\mu(\bm{k})}.
\end{align}
In the first term, the summation of $\lambda$ and $\mu$ are for the conducting (``$c$") and valence (``$v$") bands, respectively. In the second term, the range of $\lambda$ and $\mu$ are switched.

\subsubsection{Kinetic term}

The analysis of the kinetic term is essentially the same as that of the Hessian. The kinetic term features the following skew-symmetric coefficient:
\begin{align}
\Omega^{\alpha\beta}_{rr'} = -S\epsilon^{\alpha\beta}\delta_{rr'} + i(\langle \frac{\partial\Psi_G}{\partial u^\alpha_r}|\frac{\partial \Psi_G}{\partial u^\beta_{r'}}\rangle -\langle \frac{\partial\Psi_G}{\partial u^\beta_{r'}}|\frac{\partial \Psi_G}{\partial u^\alpha_r}\rangle).
\end{align}
The first order perturbation theory yields:
\begin{align}
i(\langle \frac{\partial\Psi_G}{\partial u^\alpha_r}|\frac{\partial \Psi_G}{\partial u^\beta_{r'}}\rangle -\langle \frac{\partial\Psi_G}{\partial u^\beta_{r'}}|\frac{\partial \Psi_G}{\partial u^\alpha_r}\rangle) = i\sum_{n>0}\frac{\langle\Psi_0|\frac{\partial H}{\partial u^\alpha_r}|\Psi_n\rangle\langle\Psi_n|\frac{\partial H}{\partial u^\beta_{r'}}|\Psi_0\rangle - \langle\Psi_0|\frac{\partial H}{\partial u^\beta_{r'}}|\Psi_n\rangle\langle\Psi_n|\frac{\partial H}{\partial u^\alpha_r}|\Psi_0\rangle}{(E_n-E_0)^2}
\nonumber\\
= i(JS)^2 \sum_{n>0}\frac{\langle\Psi_0|m^\alpha_r|\Psi_n\rangle\langle\Psi_n|m^\beta_{r'}|\Psi_0\rangle - \langle\Psi_0|m^\beta_{r'}|\Psi_n\rangle\langle\Psi_n|m^\alpha_r|\Psi_0\rangle}{(E_n-E_0)^2}.
\label{eq:berry}
\end{align}

In the momentum space,
\begin{align}
\Omega^{\alpha\beta}_{ll'}(\bm{q}) = -S\epsilon^{\alpha\beta} \delta_{ll'} + (JS)^2 \eta^{\alpha\beta}_{ll'}(\bm{q}),
\end{align}
where
\begin{align}
\eta^{\alpha\beta}_{ll'}(\bm{q}) = \frac{i}{N_c} \sum_{n>0}\frac{\langle\Psi_0|m^\alpha_l(\bm{q})|\Psi_n\rangle\langle\Psi_n|m^\beta_{l'}(-\bm{q})|\Psi_0\rangle - \langle\Psi_0|m^\beta_{l'}(-\bm{q})|\Psi_n\rangle\langle\Psi_n|m^\alpha_l(\bm{q})|\Psi_0\rangle}{(E_n-E_0)^2}.
\end{align}
In the diagonal basis of the equilibrium Hamiltonian $H_0$, the above expression reduces to:
\begin{align}
\eta^{\alpha\beta}_{ll'}(\bm{q}) = \frac{i}{N_c} \sum_{\bm{k}} (\sum_{\lambda \in c} \sum_{\mu\in v}-\sum_{\lambda \in v} \sum_{\mu\in c} ) \frac{\langle \bm{k}\mu|m^\alpha_l(\bm{q})|\bm{k}+\bm{q}\lambda\rangle\langle \bm{k}+\bm{q}\lambda|m^\beta_{l'}(-\bm{q})|\bm{k}\mu\rangle}{(\epsilon_\lambda(\bm{k}+\bm{q}) - \epsilon_\mu(\bm{k}))^2}.
\end{align}

\subsubsection{No direct current response \label{sec:no_current}}

In the previous section, we have pointed out that the Lagrangian formally admits a current response:
\begin{align}
\bm{J} = \frac{i}{2}\sum_{\alpha r,\beta r'} \frac{\partial}{\partial \bm{A}} (\langle \frac{\partial\Psi_G}{\partial u^\alpha_r}| \frac{\partial \Psi_G}{\partial u^\beta_{r'}}\rangle - \langle \frac{\partial\Psi_G}{\partial u^\beta_{r'}}| \frac{\partial \Psi_G}{\partial u^\alpha_{r}}\rangle) u^\alpha_r \dot{u}^\beta_{r'}.
\end{align}
We now prove that the derivative
\begin{align}
i\frac{\partial}{\partial \bm{A}} (\langle \frac{\partial\Psi_G}{\partial u^\alpha_r}| \frac{\partial \Psi_G}{\partial u^\beta_{r'}}\rangle - \langle \frac{\partial\Psi_G}{\partial u^\beta_{r'}}| \frac{\partial \Psi_G}{\partial u^\alpha_{r}}\rangle)  = 0.
\end{align}

To this end, we observe that $i(\langle \frac{\partial\Psi_G}{\partial u^\alpha_r}| \frac{\partial \Psi_G}{\partial u^\beta_{r'}}\rangle - \langle \frac{\partial\Psi_G}{\partial u^\beta_{r'}}| \frac{\partial \Psi_G}{\partial u^\alpha_{r}}\rangle)$ is a Berry curvature. We compare the Berry curvature calculated at $\bm{A}_x = \bm{A}_y= 0$ and at $\bm{A}_x = 2\pi /L_x$, $\bm{A}_y = 0$, where $L_x$ is the linear dimension of the lattice along the $x$ direction. The Hamiltonian for the later vector gauge field is related to the one for zero gauge field by a large gauge transformation:
\begin{align}
H (\bm{A}_x=\frac{2\pi}{L_x},\bm{A}_y=0) = U H(\bm{A}_x= \bm{A}_y = 0) U^\dagger,
\end{align}
where $U$ is the unitary operator that effects the gauge transformation. Therefore, the spectra of both Hamiltonians are identical, while the eigenstates are related by the gauge transformation:
\begin{align}
|\Psi_n, \bm{A}_x=\frac{2\pi}{L_x},\bm{A}_y=0 \rangle = U|\Psi_n,\bm{A}_x=\bm{A}_y = 0\rangle.
\end{align}
Meanwhile, the electron magnetic density operator is invariant under this large gauge transformation:
\begin{align}
m^\alpha_r  = U m^\alpha_r U^\dagger.
\end{align}
As a result, the matrix elements are identical:
\begin{align}
\langle \Psi_n , \bm{A}_x=\frac{2\pi}{L_x},\bm{A}_y=0 |m^\alpha_r|\Psi_0, \bm{A}_x=\frac{2\pi}{L_x},\bm{A}_y=0 \rangle = \langle \Psi_n , \bm{A}_x=\bm{A}_y=0 |m^\alpha_r|\Psi_0, \bm{A}_x=\bm{A}_y=0 \rangle.
\end{align}
It follows from Eq,~\eqref{eq:berry} that:
\begin{align}
i\left.(\langle \frac{\partial\Psi_G}{\partial u^\alpha_r}| \frac{\partial \Psi_G}{\partial u^\beta_{r'}}\rangle - \langle \frac{\partial\Psi_G}{\partial u^\beta_{r'}}| \frac{\partial \Psi_G}{\partial u^\alpha_{r}}\rangle)\right|_{\bm{A}_x=\frac{2\pi}{L_x}}=i\left.(\langle \frac{\partial\Psi_G}{\partial u^\alpha_r}| \frac{\partial \Psi_G}{\partial u^\beta_{r'}}\rangle - \langle \frac{\partial\Psi_G}{\partial u^\beta_{r'}}| \frac{\partial \Psi_G}{\partial u^\alpha_{r}}\rangle)\right|_{\bm{A}_x=0}.
\end{align}
Taking the limit $L_x\to\infty$, we obtain:
\begin{align}
i\frac{\partial}{\partial \bm{A}_x} (\langle \frac{\partial\Psi_G}{\partial u^\alpha_r}| \frac{\partial \Psi_G}{\partial u^\beta_{r'}}\rangle - \langle \frac{\partial\Psi_G}{\partial u^\beta_{r'}}| \frac{\partial \Psi_G}{\partial u^\alpha_{r}}\rangle)  = 0.
\end{align}
By the same token, the derivative with respect to $\bm{A}_y$ vanishes as well.

\subsubsection{Linear magnetoelectric coefficient}

The linear magnetoelectric coefficient $\Upsilon$ is the Berry curvature between $\bm{A}$ and $u^\alpha_r$:
\begin{align}
\bm{\Upsilon}^{\alpha}_{r}  = i (\langle\frac{\partial\Psi_G}{\partial \bm{A}}|\frac{\partial\Psi_G}{\partial u^\alpha_r}\rangle - \langle\frac{\partial\Psi_G}{\partial u^\alpha_r}|\frac{\partial\Psi_G}{\partial \bm{A}}\rangle).
\end{align}
First order perturbation theory yields:
\begin{align}
\bm{\Upsilon}^{\alpha}_{r} = i\sum_{n>0}\frac{\langle\Psi_0|\frac{\partial H}{\partial \bm{A}}|\Psi_n\rangle\langle\Psi_n|\frac{\partial H}{\partial u^\alpha_r}|\Psi_0\rangle - \langle\Psi_0|\frac{\partial H}{\partial u^\alpha_r}|\Psi_n\rangle\langle\Psi_n|\frac{\partial H}{\partial \bm{A}}|\Psi_0\rangle}{(E_n-E_0)^2}
\nonumber\\
= ie(JS)\sum_{n>0}\frac{\langle\Psi_0|\bm{I}|\Psi_n\rangle\langle\Psi_n|m^\alpha_r|\Psi_0\rangle - \langle\Psi_0|m^\alpha_r|\Psi_n\rangle\langle\Psi_n|\bm{I}|\Psi_0\rangle}{(E_n-E_0)^2}.
\end{align}
Here, the derivative:
\begin{align}
\frac{\partial H}{\partial \bm{A}} = -e\bm{I} = -e\sum_{\bm{k}s} \nabla\epsilon_{\bm{k}} c^\dagger_{\bm{k}s} c^{\phantom\dagger}_{\bm{k}s},
\end{align}
is the electron current operator up to a minus sign.

Owing to the translation symmetry, $\bm{\Upsilon}^{\alpha}_r$ depends only on the sublattice label of the site $r$. We therefore define:
\begin{align}
\bm{\Upsilon}^{\alpha}_l = \frac{1}{N_c}\sum_{r\in l} \bm{\Upsilon}^{\alpha}_r = \frac{ie(JS)}{N_c}\sum_{n>0}\frac{\langle\Psi_0|\bm{I}|\Psi_n\rangle\langle\Psi_n|m^\alpha_l(\bm{q}=0)|\Psi_0\rangle - \langle\Psi_0|m^\alpha_l(\bm{q}=0)|\Psi_n\rangle\langle\Psi_n|\bm{I}|\Psi_0\rangle}{(E_n-E_0)^2}.
\end{align}
In the diagonal basis of the equilibrium Hamiltonian,
\begin{align}
\bm{\Upsilon}^{\alpha}_l = \frac{ie(JS)}{N_c}\sum_{\bm{k}} (\sum_{\lambda\in c} \sum_{\mu\in v}-\sum_{\lambda\in v} \sum_{\mu\in c}) \frac{\langle \bm{k}\mu|\bm{I}|\bm{k}\lambda\rangle\langle \bm{k}\lambda|m^\alpha_l(\bm{q}=0)|\bm{k}\mu\rangle}{(\epsilon_\lambda(\bm{k}) - \epsilon_\mu(\bm{k}))^2}.
\end{align}

In fact, $\Upsilon = 0$ by symmetry in the cases we are interested in. We may verify this fact by direct calculation.

\subsubsection{Nonlinear magnetoelectric coefficient}

The nonlinear magnetoelectric coefficient $\bm{Z}$ is more involved. Using the \emph{second order} Rayleigh-Schr\"{o}dinger perturbation theory, we find:
\begin{align}
\bm{Z}^{\alpha\beta}_{rr'} = \frac{1}{2}(\bm{F}^{\alpha\beta}_{rr'} + \bm{F}^{\beta\alpha}_{r' r}).
\end{align}
where
\begin{align}
\bm{F}^{\alpha\beta}_{rr'} = ie(JS)^2 \sum_{m\neq 0}\sum_{n\neq 0}\frac{(m^\beta_{r'}, \bm{I}, m^\alpha_r) + (\bm{I}, m^\beta_{r'}, m^\alpha_r) - (m^\alpha_r, m^\beta_{r'}, \bm{I}) - (m^\beta_{r'}, m^\alpha_r, \bm{I})}{(0,n,m^2)}
\nonumber\\
+ ie(JS)^2 \sum_{m\neq 0} \sum_{n \neq 0} \frac{(\bm{I}, m^\beta_{r'}, m^\alpha_r)+(\bm{I}, m^\alpha_r, m^\beta_{r'}) - (m^\alpha_r, \bm{I}, m^\beta_{r'}) - (m^\alpha_r, m^\beta_{r'}, \bm{I})}{(0,n^2,m)}
\nonumber\\
-ie(JS)\delta_{rr'}\delta_{\alpha\beta}\sum_{n\neq k}\frac{(\bm{I}, m^{\snum3}_r) - (m^{\snum3}_r, \bm{I}) }{(k,n^2)}.
\end{align}
We have used the short hand notations:
\begin{align}
\frac{(O_1,O_2,O_3)}{(k,n,m^2)} = \frac{\langle k|O_1|n\rangle \langle n|O_2|m\rangle \langle m|O_3|k\rangle}{(E_n-E_k)(E_m-E_k)^2};
\quad
\frac{(O_1,O_2,O_3)}{(k,n^2,m)} = \frac{\langle k|O_1|n\rangle \langle n|O_2|m\rangle \langle m|O_3|k\rangle}{(E_n-E_k)^2(E_m-E_k)};
\nonumber\\
\frac{(O_1,O_2)}{(k,n^2)} = \frac{\langle k|O_1|n\rangle \langle n|O_2|k\rangle}{(E_n-E_k)^2}.
\end{align}

In the momentum space,
\begin{align}
\bm{Z}^{\alpha\beta}_{ll'}(\bm{q}) =  \frac{1}{2} ( \bm{F}^{\alpha\beta}_{ll'}(\bm{q}) + \bm{F}^{\beta\alpha}_{l'l}(\bm{q})^\ast).
\end{align}
which shows explicitly that $\bm{Z}(\bm{q})$ is an Hermitian matrix. $\bm{F}^{\alpha\beta}_{ll'}(\bm{q})$ is given by:
\begin{align}
\bm{F}^{\alpha\beta}_{ll'}(\bm{q}) = e(JS)^2 \bm{f}^{\alpha\beta}_{ll'}(\bm{q}) - e(JS) \bm{g}_l\delta_{\alpha\beta}\delta_{ll'},
\end{align}
where
\begin{align}
\bm{f}^{\alpha\beta}_{ll'}(\bm{q}) = \frac{i}{N_c}\sum_{m,n\neq0}\frac{(m^\beta_{l'}(-\bm{q}), \bm{I}, m^\alpha_l(\bm{q})) + (\bm{I}, m^\beta_{l'}(-\bm{q}), m^\alpha_l(\bm{q})) - (m^\beta_{l'}(-\bm{q}), m^\alpha_l(\bm{q}), \bm{I}) - (m^\alpha_l(\bm{q}), m^\beta_{l'}(-\bm{q}), \bm{I})}{(0,n,m^2)} 
\nonumber\\
+ \frac{(\bm{I},m^\alpha_l(\bm{q}),m^\beta_{l'}(-\bm{q})) + (\bm{I}, m^\beta_{l'}(-\bm{q}), m^\alpha_{l}(\bm{q})) - (m^\alpha_{l}(\bm{q}), \bm{I}, m^\beta_{l'}(-\bm{q})) - (m^\alpha_l(\bm{q}), m^\beta_{l'}(-\bm{q}), \bm{I}) }{(0,n^2,m)}.
\end{align}
and
\begin{align}
\bm{g}_l = \frac{i}{N_c}\sum_{n\neq 0}\sum_{n\neq k}\frac{(\bm{I}, m^{\snum3}_l(\bm{q}=0)) - (m^{\snum3}_l(\bm{q}=0), \bm{I}) }{(k,n^2)}.
\end{align}

In the next step, we switch to the diagonal basis of the electron Hamiltonian at equilibrium. Computing the quantity $\bm{g}_l$ is straightforward:
\begin{align}
\bm{g}_l = \frac{i}{N_c}\sum_{\bm{k}} (\sum_{\lambda\in c}\sum_{\mu \in v} - \sum_{\lambda \in v}\sum_{\mu \in c}) \frac{\langle \bm{k}\mu|\bm{I}|\bm{k}\lambda\rangle \langle \bm{k}\lambda|m^{\snum3}_l(\bm{q}=0)|\bm{k}\mu\rangle}{(\epsilon_\lambda(\bm{k}) - \epsilon_\mu(\bm{k}))^2}.
\end{align}
The expression for $\bm{f}^{\alpha\beta}_{ll'}$ is significantly more complex:
\begin{subequations}
\begin{align}
\bm{f}^{\alpha\beta}_{ll'}(\bm{q}) =  \frac{i}{N_c}\sum_{\bm{k}} (\sum_{\alpha \in v}\sum_{\beta,\gamma \in c}-\sum_{\alpha \in c}\sum_{\beta,\gamma \in v})\, \mathrm{Summand},
\end{align}
with
\begin{align}
\mathrm{Summand} = \frac{(m^\beta_{l'}(-\bm{q}), \bm{I}, m^\alpha_l(\bm{q}))}{(\bm{k}+\bm{q}\alpha,\bm{k}\beta,\bm{k}\gamma^2)} 
+ \frac{(\bm{I},m^\beta_{l'}(-\bm{q}),m^\alpha_l(\bm{q}))}{(\bm{k}+\bm{q}\alpha,\bm{k}+\bm{q}\beta,\bm{k}\gamma^2)}
-\frac{(m^\alpha_l(\bm{q}),m^\beta_{l'}(-\bm{q}),\bm{I})}{(\bm{k}\alpha,\bm{k}+\bm{q}\beta,\bm{k}\gamma^2)}
-\frac{(m^\beta_{l'}(-\bm{q}),m^\alpha_l(\bm{q}),\bm{I})}{(\bm{k}+\bm{q}\alpha,\bm{k}\beta,\bm{k}+\bm{q}\gamma^2)}
\nonumber\\
-\frac{(m^\alpha_l(\bm{q}), \bm{I}, m^\beta_{l'}(-\bm{q}))}{(\bm{k}\alpha,\bm{k}+\bm{q}\beta^2,\bm{k}+\bm{q}\gamma)} 
- \frac{(m^\alpha_l(\bm{q}),m^\beta_{l'}(-\bm{q}),\bm{I})}{(\bm{k}\alpha,\bm{k}+\bm{q}\beta^2,\bm{k}\gamma)}
+ \frac{(\bm{I},m^\beta_{l'}(-\bm{q}),m^\alpha_l(\bm{q}))}{(\bm{k}+\bm{q}\alpha,\bm{k}+\bm{q}\beta^2,\bm{k}\gamma)}
+\frac{(\bm{I},m^\alpha_l(\bm{q}),m^\beta_{l'}(-\bm{q}))}{(\bm{k}\alpha,\bm{k}\beta^2,\bm{k}+\bm{q}\gamma)}.
\end{align}
\end{subequations}

\subsection{Application to hexagonal tetrahedral order}

We consider the hexagonal lattice tetrahedral state:
\begin{align}
\hat{n}_r = \frac{1}{\sqrt{3}} (\hat{e}^{\snum1} e^{i\bm{Q}_1\cdot \bm{r}} + \hat{e}^{\snum2} e^{i\bm{Q}_2\cdot \bm{r}} + \hat{e}^{\snum3} e^{i\bm{Q}_3\cdot \bm{r}}),
\end{align}
where the notations are the same as that of Sec.~\ref{sec:hexagonal}. We define a set of local spin frames:
\begin{align}
\hat{e}^{\snum1}_r &= -\frac{2}{\sqrt{6}} \hat{e}^{\snum1} e^{i\bm{Q}_1\cdot \bm{r}} + \frac{1}{\sqrt{6}} \hat{e}^{\snum2} e^{i\bm{Q}_2\cdot \bm{r}} +  \frac{1}{\sqrt{6}} \hat{e}^{\snum3} e^{i\bm{Q}_3\cdot \bm{r}}.
\nonumber\\
\hat{e}^{\snum2}_r &=  -\frac{1}{\sqrt{2}} \hat{e}^{\snum2} e^{i\bm{Q}_2\cdot \bm{r}} + \frac{1}{\sqrt{2}} \hat{e}^{\snum3} e^{i\bm{Q}_3\cdot \bm{r}}.
\nonumber\\
\hat{e}^{\snum3}_r &= \frac{1}{\sqrt{3}} \hat{e}^{\snum1} e^{i\bm{Q}_1\cdot \bm{r}} + \frac{1}{\sqrt{3}} \hat{e}^{\snum2} e^{i\bm{Q}_2\cdot \bm{r}} +  \frac{1}{\sqrt{3}} \hat{e}^{\snum3} e^{i\bm{Q}_3\cdot \bm{r}}.
\end{align}
This specific choice of the local frames respects the SSG translation symmetries; as a direct consequence, all of the Lagrangian parameters, $\Omega$, $\Phi$, and $Z$ features the translation symmetry of the underlying hexagonal lattice. It is then possible to present the dispersion relation of the spin waves in the \emph{lattice} Brillouin zone rather than the magnetic Brillouin zone.

The electron Hamiltonian at equilibrium reads:
\begin{align}
H = \sum_{\bm{k}\in \mathrm{MBZ}} \overline{\Psi}(\bm{k}) H(\bm{k}) \Psi(\bm{k}) = \sum_{\bm{k}\in \mathrm{MBZ}} \overline{\Psi}(\bm{k})\begin{pmatrix}
\epsilon_{\bm{k}} & -\frac{JS}{\sqrt{3}}\sigma_1 & -\frac{JS}{\sqrt{3}} \sigma_2 & -\frac{JS}{\sqrt{3}} \sigma_3 \\
-\frac{JS}{\sqrt{3}}\sigma_1 & \epsilon_{\bm{k}+\bm{Q}_1} & -\frac{JS}{\sqrt{3}}\sigma_3 & -\frac{JS}{\sqrt{3}}\sigma_2 \\
-\frac{JS}{\sqrt{3}}\sigma_2 & -\frac{JS}{\sqrt{3}}\sigma_3 & \epsilon_{\bm{k}+\bm{Q}_2} & -\frac{JS}{\sqrt{3}}\sigma_1\\
-\frac{JS}{\sqrt{3}}\sigma_3 & -\frac{JS}{\sqrt{3}}\sigma_2 & -\frac{JS}{\sqrt{3}}\sigma_1 & \epsilon_{\bm{k}+\bm{Q}_3}
\end{pmatrix} \Psi(\bm{k}).
\end{align}
The summation is over the magnetic Brillouin zone. $\Psi(\bm{k}) = (c_{\uparrow}(\bm{k}),c_{\downarrow}(\bm{k}),c_{\uparrow}(\bm{k}+\bm{Q}_1),c_{\downarrow}(\bm{k}+\bm{Q}_1),c_{\uparrow}(\bm{k}+\bm{Q}_2),c_{\downarrow}(\bm{k}+\bm{Q}_2),c_{\uparrow}(\bm{k}+\bm{Q}_3),c_{\downarrow}(\bm{k}+\bm{Q}_3))^T$ is a column vector.
\begin{align}
\epsilon_{\bm{k}} = -2t [\cos(\bm{k} \cdot \bm{a}_1)+\cos(\bm{k} \cdot \bm{a}_2)+\cos(\bm{k} \cdot \bm{a}_3)],
\end{align}
is the tight binding dispersion on hexagonal lattice. $\bm{a}_1 = (1,0)$, $\bm{a}_2 = (-\frac{1}{2},\frac{\sqrt{3}}{2})$, and $\bm{a}_3 = (-\frac{1}{2},-\frac{\sqrt{3}}{2})$. We may diagonalize the Hamiltonian by a unitary transformation:
\begin{align}
H = \sum_{\bm{k}\in \mathrm{MBZ}} \overline{\Psi}(\bm{k}) U(\bm{k}) E(\bm{k}) U(\bm{k})^\dagger \Psi(\bm{k}) = \sum_{\bm{k} \in \mathrm{MBZ}} \sum_\lambda \epsilon_{\lambda}(\bm{k}) c^\dagger_{\lambda}(\bm{k})c^{\phantom\dagger}_{\lambda}(\bm{k}),
\end{align}
where $U(\bm{k})$ is the unitary matrix that diagonalizes $H(\bm{k})$. $\lambda$ labels the eigenvalues or Bloch bands.

The electric current operator $\bm{I}$ is given by:
\begin{align}
\bm{I}_{x,y} = \sum_{\bm{k}\in \mathrm{MBZ}} \overline{\Psi}(\bm{k})V_{x,y}(\bm{k})\Psi(\bm{k}) =  \sum_{\bm{k}\in \mathrm{MBZ}} \sum_{\lambda \mu} [U(\bm{k})^\dagger V_{x,y}(\bm{k}) U(\bm{k})]_{\lambda\mu} c^\dagger_{\lambda}(\bm{k})c^{\phantom\dagger}_{\mu}(\bm{k}).
\end{align}
In the second equality, we have switched to the diagonal basis of $H$. The matrices:
\begin{align}
V_{x,y}(\bm{k}) = \mathrm{diag}(v_{x,y}(\bm{k}), v_{x,y}(\bm{k}+\bm{Q}_1), v_{x,y}(\bm{k}+\bm{Q}_2), v_{x,y}(\bm{k}+\bm{Q}_3)),
\end{align}
with
\begin{align}
v_x(\bm{k}) = 2t(\sin(\bm{k}\cdot \bm{a}_1) - \frac{1}{2}\sin(\bm{k} \cdot \bm{a}_2) - \frac{1}{2}\sin(\bm{k} \cdot \bm{a}_3));
\,
v_y(\bm{k}) = 2t(\frac{\sqrt{3}}{2}\sin(\bm{k} \cdot \bm{a}_2) - \frac{\sqrt{3}}{2}\sin(\bm{k} \cdot \bm{a}_3));
\end{align}

The electron magnetic moment operator:
\begin{align}
m^\alpha(\bm{q}) = \sum_{\bm{k}\in \mathrm{MBZ}} \overline{\Psi}(\bm{k}) O^\alpha \Psi(\bm{k}+\bm{q}) = \sum_{\bm{k}\in \mathrm{MBZ}} \sum_{\lambda \mu} [U(\bm{k})^\dagger O^\alpha U(\bm{k}+\bm{q})]_{\lambda\mu} c^\dagger_{\lambda}(\bm{k})c^{\phantom\dagger}_{\mu}(\bm{k}+\bm{q}).
\end{align}
The matrices $O^{\snum1}$, etc. are given by:
\begin{align}
O^{\snum1} =  \begin{pmatrix}
0 & -\frac{2}{\sqrt{6}}\sigma^{\snum1} & \frac{1}{\sqrt{6}} \sigma^{\snum2} & \frac{1}{\sqrt{6}} \sigma^{\snum3} \\
-\frac{2}{\sqrt{6}}\sigma^{\snum1} & 0 & \frac{1}{\sqrt{6}}\sigma^{\snum3} & \frac{1}{\sqrt{6}}\sigma^{\snum2} \\
\frac{1}{\sqrt{6}}\sigma^{\snum2} & \frac{1}{\sqrt{6}}\sigma^{\snum3} & 0 & -\frac{2}{\sqrt{6}}\sigma^{\snum1} \\
\frac{1}{\sqrt{6}}\sigma^{\snum3} & \frac{1}{\sqrt{6}}\sigma^{\snum2} & -\frac{2}{\sqrt{6}}\sigma^{\snum1} & 0
\end{pmatrix};
\quad
O^{\snum2} =  \begin{pmatrix}
0 & 0 & -\frac{1}{\sqrt{2}} \sigma^{\snum2} & \frac{1}{\sqrt{2}} \sigma^{\snum3} \\
0 & 0 & \frac{1}{\sqrt{2}}\sigma^{\snum3} & -\frac{1}{\sqrt{2}}\sigma^{\snum2} \\
-\frac{1}{\sqrt{2}}\sigma^{\snum2} & \frac{1}{\sqrt{2}}\sigma^{\snum3} & 0 & 0 \\
\frac{1}{\sqrt{2}}\sigma^{\snum3} & -\frac{1}{\sqrt{2}}\sigma^{\snum2} & 0 & 0
\end{pmatrix};
\nonumber\\
O^{\snum3} =  \begin{pmatrix}
0 & \frac{1}{\sqrt{3}}\sigma^{\snum1} & \frac{1}{\sqrt{3}} \sigma^{\snum2} & \frac{1}{\sqrt{3}} \sigma^{\snum3} \\
\frac{1}{\sqrt{3}}\sigma^{\snum1} & 0 & \frac{1}{\sqrt{3}}\sigma^{\snum3} & \frac{1}{\sqrt{3}}\sigma^{\snum2} \\
\frac{1}{\sqrt{3}}\sigma^{\snum2} & \frac{1}{\sqrt{3}}\sigma^{\snum3} & 0 & \frac{1}{\sqrt{3}}\sigma^{\snum1}\\
\frac{1}{\sqrt{3}}\sigma^{\snum3} & \frac{1}{\sqrt{3}}\sigma^{\snum2} & \frac{1}{\sqrt{3}}\sigma^{\snum1} & 0
\end{pmatrix}.
\end{align}
Using these, we may calculate the Lagrangian parameters following the recipe of Sec.~\ref{sec:parameters}.

We find by explicit calculation that the linear magnetoelectric coefficient $\bm{\Upsilon}^{\alpha}_l = 0$. This is consistent with the symmetry: the magnetic order features a $C^z_2$ rotational symmetry about a site, and the $\mathbf{q}=0$ deformations preserve this symmetry. This symmetry forbids coupling to a uniform electric field. Furthermore, by a direct calculation, we have verified that:
\begin{align}
i\frac{\partial}{\partial \bm{A}} (\langle \frac{\partial\Psi_G}{\partial u^\alpha_r}| \frac{\partial \Psi_G}{\partial u^\beta_{r'}}\rangle - \langle \frac{\partial\Psi_G}{\partial u^\beta_{r'}}| \frac{\partial \Psi_G}{\partial u^\alpha_{r}}\rangle)  \to 0,
\end{align}
as the system size increases, thereby verifying the result of Sec.~\ref{sec:no_current}.

\end{document}